\documentclass[11pt,letterpaper,twoside]{article}
\usepackage{color,graphicx,natbib,lscape,here,geometry,setspace,multirow,enumitem}
\usepackage[dvipsnames]{xcolor}
\usepackage{graphicx}
\usepackage{authblk}
\usepackage{silence}
\usepackage{rotating}
\usepackage{multirow}

\usepackage{booktabs}
\usepackage{color}
\usepackage{setspace}
\usepackage{algorithm2e}
\usepackage{enumitem}
\usepackage{tikz}

\usepackage{newtxtext,newtxmath}
\newcommand*{\scfont}{\fontfamily{ptm}\selectfont}

\definecolor{nblue}{HTML}{000660}
\usepackage[colorlinks=true,urlcolor=nblue,linkcolor=nblue,citecolor=nblue]{hyperref}
\usepackage{bigstrut}

\usepackage{tabularx}
\usepackage{threeparttable}
\usepackage{dcolumn}

\usepackage{etoolbox} 
\makeatletter
\patchcmd{\BR@backref}{\newblock}{\newblock[}{}{}
\patchcmd{\BR@backref}{\par}{]\par}{}{}
\makeatother

\usepackage{pdflscape}
\usepackage{afterpage}

\usepackage{longtable}
\usepackage{array}
\newcolumntype{C}[1]{>{\centering\arraybackslash}p{#1}}

\usepackage{comment}
\usepackage{mathtools}

\usepackage[hang,flushmargin]{footmisc} 
\usepackage[british]{babel}

\pdfoutput=1
 \usepackage{float}
 \restylefloat{table}

\usepackage[title,titletoc]{appendix}
\makeatletter

\renewenvironment{appendices}{%
    \begin{oldappendices}%
    \renewcommand{\thefigure}{\ifnum \c@section>\z@ \thesection.\fi\@arabic\c@figure}%
    \@addtoreset{figure}{section}%
    \renewcommand{\thetable}{\ifnum \c@section>\z@ \thesection.\fi\@arabic\c@table}%
    \@addtoreset{table}{section}}{%
    \end{oldappendices}%
}\makeatother

\usepackage{titlesec} 
\titleformat{\section}[block]{\large}{\thesection. }{0em}{\MakeUppercase} 
\titleformat{\subsection}[block]{\large}{\thesubsection. }{0em}{\itshape} 
\titleformat{\subsubsection}[block]{\large}{}{0em}{\itshape} 

\let\natbibcitet\citet
\renewcommand\citet{\bibpunct{(}{)}{,}{a}{,}{,}\natbibcitet}
\let\natbibcitep\citep
\renewcommand\citep{\bibpunct{(}{)}{;}{a}{,}{;}\natbibcitep}
\newcommand{\bi}{\begin{itemize}}
\newcommand{\ei}{\end{itemize}}
\newcommand{\be}{\begin{equation}}
\newcommand{\ee}{\end{equation}}
\defcitealias{ieo14}{IEO, 2014}

\makeatletter
\def\ubar#1{\underline{\sbox\tw@{$#1$}\dp\tw@\z@\box\tw@}}
\def\obar#1{\overline{\sbox\tw@{$#1$}\dp\tw@\z@\box\tw@}}
\makeatother

\usepackage{bm}
\usepackage{caption}
\usepackage{subcaption}

\makeatletter\let\p@subfigure\thefigure\makeatother

\floatstyle{plaintop}
\restylefloat{table}

\usepackage{fancyhdr} 
\usepackage{cleveref}
\crefname{chapter}{Chapter}{Chapters}
\crefname{section}{Section}{Sections}
\crefname{subsection}{Section}{Sections}
\crefname{subsubsection}{Section}{Sections}
\crefname{figure}{Figure}{Figures}
\crefname{table}{Table}{Tables}
\crefname{equation}{Equation}{Equations}
\crefname{appendix}{Appendix}{Appendices}
\crefname{appendices}{Appendix}{Appendices}

\crefname{appsec}{Appendix}{Appendices}

\usepackage{catoptions}
\makeatletter

\def\Autoref#1{%
  \begingroup
  \edef\reserved@a{\cpttrimspaces{#1}}%
  \ifcsndefTF{r@#1}{%
    \xaftercsname{\expandafter\testreftype\@fourthoffive}
      {r@\reserved@a}.\\{#1}%
  }{%
    \ref{#1}%
  }%
  \endgroup
}
\def\testreftype#1.#2\\#3{%
  \ifcsndefTF{#1autorefname}{%
    \def\reserved@a##1##2\@nil{%
      \uppercase{\def\ref@name{##1}}%
      \csn@edef{#1autorefname}{\ref@name##2}%
      \autoref{#3}%
    }%
    \reserved@a#1\@nil
  }{%
    \autoref{#3}%
  }%
}
\makeatother

\newcolumntype{d}[1]{D{.}{.}{#1}}

\title{\LARGE{A multi-country dynamic factor model with stochastic volatility for euro area business cycle analysis}}
\author{\large{\uppercase{Florian Huber}$^{1}$, \uppercase{Michael Pfarrhofer}$^{1,2}$ and \uppercase{Philipp Piribauer}$^{3}$\thanks{\textit{Corresponding author}: Philipp Piribauer, Austrian Institute of Economic Research (WIFO), Arsenal 20, 1030 Vienna, Austria. \textit{E-mail}: \href{mailto:philipp.piribauer@wifo.ac.at}{philipp.piribauer@wifo.ac.at}. The research carried out in this paper was supported by funds of the Oesterreichische Nationalbank (Jubilaeumsfond project number: 17382). \textit{Date}: \today.}}\\
\vspace*{-0.5em}\normalsize{$^1$\textit{University of Salzburg, Salzburg Centre of European Union Studies}\\$^2$\textit{Vienna University of Economics and Business}\\$^3$\textit{Austrian Institute of Economic Research}}}
\date{}


\def\equationautorefname~#1\null{%
  Eq.~(#1)\null
}
\def\equationautorefname~#1\null{
Eq.~(#1)\null
}

\begin{document}
\maketitle\thispagestyle{empty}\normalsize\vspace*{-2em}\small

\begin{center}
\begin{minipage}{0.8\textwidth}
\noindent\small This paper develops a dynamic factor model that uses euro area (EA) country-specific information on output and inflation to estimate an area-wide measure of the output gap. Our model assumes that output and inflation can be decomposed into country-specific stochastic trends and a common cyclical component. Comovement in the trends is introduced by imposing a factor structure on the shocks to the latent states. We moreover introduce flexible stochastic volatility specifications to control for  heteroscedasticity in the measurement errors and innovations to the latent states. Carefully specified shrinkage priors allow for pushing the model towards a homoscedastic specification, if supported by the data.  Our measure of the output gap closely tracks other commonly adopted measures, with small differences in magnitudes and timing. To assess whether the model-based output gap helps in forecasting inflation, we perform an out-of-sample forecasting exercise. The findings indicate that our approach yields superior inflation forecasts, both in terms of point and density predictions.
\MakeUppercase{\textit{keywords}}: European business cycles, dynamic factor model, factor stochastic volatility, inflation forecasting\\
\end{minipage}
\end{center}
\bigskip\normalsize
\renewcommand{\thepage}{\arabic{page}}
\setcounter{page}{1}
\newpage
\section{Introduction}\label{sec:intro}
Effective policy making in central banks such as the European Central Bank (ECB)  requires accurate measures of latent quantities such as the output gap to forecast key quantities of interest like inflation across euro area (EA) member states. Since using aggregate EA data potentially masks important country-specific dynamics, exploiting country-level information could help in obtaining more reliable estimates of the output gap that is consequently used in Phillips curve-type models to forecast inflation.

In this paper, we exploit cross-sectional information on output and inflation dynamics to construct a multi-country model for the EA. The proposed framework  aims to combine the literature on output gap modeling \citep[see, among many others,][]{kuttner1994estimating, orphanides2002unreliability, basistha2007new, planas2008bayesian} that focuses on estimating the output gap based on data for a single country/regional aggregate, the literature on dynamic factor models \citep{otrok1998bayesian, kim1999state, koseotrokwhiteman, breitungeickmeierEL, jarocinski2018inflation} and the literature on inflation forecasting \citep{stock1999forecasting,stock2007has}. 

Our model assumes that country-specific business cycles are driven by a common latent factor, effectively  exploiting cross-sectional information in the data. Moreover, we assume that output and inflation feature a non-stationary country-specific component. To control for potential comovement in these trend terms, we assume that the corresponding shocks to the states feature a factor structure. The resulting factor model features stochastic volatility (SV) in the spirit of \cite{aguilar2000bayesian} and thus provides a parsimonious way of controlling for heteroscedasticity. Since successful forecasting models typically allow for SV \citep{clark2011realtime,clarkravazzolo2015JAE,HUBER2016818,HuberFeldkircher2019JBES}, we also allow for time-variation in the error variances across the remaining state innovations and the measurement errors. One methodological key innovation is the introduction of global-local shrinkage priors on the error variances of the state equations describing the law of motion of the logarithmic volatility components, effectively shrinking the system towards a homoscedastic specification, if applicable. 

This increased flexibility, however, is costly in terms of additional parameters to estimate. We thus follow the recent literature on state space modeling  \citep{fruhwirth2010stochastic, belmonte2014hierarchical, kastner2014ancillarity, feldkircher2017sophisticated, bitto2016achieving} and exploit a non-centered parameterization of the  model \citep[see][]{fruhwirth2010stochastic} to test whether SV is supported by the data. The non-centered parameterization allows treating the square root of the process innovation variances as standard regression coefficients, implying that conventional shrinkage priors can be used. Here we follow \cite{griffin2010inference} and use a variant of the Normal-Gamma (NG) shrinkage prior that introduces a global shrinkage component that applies to all process variances simultaneously, forcing them towards zero. Local shrinkage parameters are then used to drag sufficient posterior mass away from zero even in the presence of strong global shrinkage, allowing for non-zero process variances if required.

When applied to data for ten EA countries over the time period from 1997:Q1 to 2018:Q4, we find that our output gap measure closely tracks other measures reported in previous studies  \citep{planas2008bayesian, jarocinski2018inflation} as well as gaps obtained by utilizing standard tools commonly used in policy institutions. We moreover perform historical decompositions to gauge the importance of area-wide as opposed to country-specific shocks for describing inflation movements. These measures reveal that inflation is strongly driven by common business cycle dynamics, underlining the importance of controlling for a common business cycle. We then turn to assessing whether there exists a Phillips curve across EA countries by simulating a negative one standard deviation business cycle shock. This exercise points towards a robust relationship between the common gap component and inflation, with magnitudes differing across countries.

The main part of the empirical application applies  our modeling approach to forecast inflation,  paying particular attention on whether the inclusion of a common output gap improves predictive capabilities. Since inflation across countries is driven by a term measuring trend inflation and the output gap, our framework can be interpreted as a New Keynesian Phillips curve, akin to \cite{stella2013state}.  Compared to a set of simpler alternatives that range from univariate benchmark models to models that use alternative ways to calculate the output gap, the proposed model yields more precise point and density forecasts for inflation. 

The remainder of the paper is structured as follows. Section 2 describes the econometric framework. After providing an overview of the model, we discuss the Bayesian prior choice and briefly summarize the main steps involved in estimating the model. Section 3 presents the empirical application, starting with a summary of the dataset and inspects various key features of our model. The section moreover studies the dynamic impact of business cycle shocks to the country-specific output and inflation series. In a forecasting exercise, Section 4 compares the out-of-sample predictive performance of our model with other specifications. The final section summarizes and concludes the paper.

\section{Econometric framework}\label{sec: econometrics}
\subsection{A dynamic factor model for the euro area}
In this section we describe the framework to estimate the euro area output gap using disaggregate country-level information. Let $y_{it}$ and $\pi_{it}$ denote output and inflation for country $i=1,\dots,N$ in period $t=1,\dots,T$, respectively. For notational simplicity, we define $k \in \{y, \pi \}$.

Country-specific output and inflation are driven by unobserved common non-stationary trend components $\tau_{kit}$ that aim to capture low-frequency movements, while a common cyclical component $g_t$ tracks mid- to high-frequency fluctuations in inflation and output. These unobserved (latent) quantities are related to the observed quantities through a set of measurement equations:
\begin{align}
y_{it} &= \tau_{yit}+\alpha_{i}g_{t}+\epsilon_{yit}, \label{eq: obs_output}\\
{\pi}_{it} &= \tau_{\pi it}+\beta_{i}g_{t}+\epsilon_{\pi it},\label{eq: obs_pi}\\
\epsilon_{kit} & \sim\mathcal{N}(0,\text{e}^{h_{kit}}).\label{eq: measurement-err}
\end{align}
These equations imply that the trend components can loosely be interpreted as country-specific trend inflation and potential output for the $i$th country, respectively. Moreover, the stationary component of output and inflation depends on the common cycle $g_t$ through a set of idiosyncratic factor loadings $\alpha_i$ and $\beta_i$ and measurement errors that feature time-varying variances $e^{h_{kit}}$. It is worth stressing that \autoref{eq: obs_pi} represents a country-specific Phillips curve that establishes a relationship between inflation and the area-wide output gap $g_t$. One key goal of this paper is to assess whether there exists a Phillips curve across EA countries by inspecting $\beta_i$ and functions thereof.

The country-specific components in \autoref{eq: obs_output} and \autoref{eq: obs_pi} are stacked in $\bm{\tau}_{yt} = (\tau_{y1t},\hdots,\tau_{yNt})'$ and $\bm{\tau}_{\pi t} = (\tau_{\pi 1t},\hdots,\tau_{\pi Nt})'$ and evolve according to a VAR(2) process given by the state equation
\begin{equation}
\underbrace{\begin{bmatrix}
\bm{\tau}_{yt}\\ \bm{\tau}_{\pi t} \\ g_t 
\end{bmatrix}}_{\bm{f}_t}
 =
 \underbrace{\begin{bmatrix}
\bm{I}_N & \hdots & \bm{0} \\
\vdots & \bm{I}_N & \vdots \\
\bm{0} & \hdots & \phi_1 \\
\end{bmatrix}
}_{\bm{\Phi}_1}
\underbrace{
\begin{bmatrix}
\bm{\tau}_{yt-1}\\ \bm{\tau}_{\pi t-1} \\ g_{t-1} 
\end{bmatrix}
}_{\bm{f}_{t-1}} +
\underbrace{\begin{bmatrix}
0 & \hdots & 0 \\
\vdots & \ddots & \vdots \\
0 & \hdots & \phi_2 \\
\end{bmatrix}
}_{\bm{\Phi}_2}
\underbrace{\begin{bmatrix}
\bm{\tau}_{yt-2}\\ \bm{\tau}_{\pi t-2} \\ g_{t-2} 
\end{bmatrix}}_{\bm{f}_{t-2}}
+ \underbrace{\begin{bmatrix} \bm{\eta}_{yt} \\ \bm{\eta}_{\pi t} \\ \eta_{gt}\end{bmatrix}}_{\bm{\eta}_t}. \label{eq: stateeq}
\end{equation}
By defining  $\bm{\Phi}=( \bm{\Phi}_1, \bm{\Phi}_2)$ and $\bm{F}_t = (\bm{f}'_{t-1}, \bm{f}'_{t-2})'$, \autoref{eq: stateeq} can be written more compactly as
\begin{equation*}
\bm{f}_t = \bm{\Phi} \bm{F}_t + \bm{\eta}_t, \quad \bm{\eta}_t \sim \mathcal{N}(\bm{0},\bm{\Sigma}_t).
\end{equation*} 
Here, $\bm{\eta}_t$ denotes the stacked error terms that follows a multivariate Gaussian distribution with zero mean and time-varying variance covariance matrix $\bm{\Sigma}_t$ that is specified below.

For the unrestricted AR(2) parameters $\phi_1$ and $\phi_2$ we follow \citet{planas2008bayesian} and reparameterize the state equation coefficients of $g_t$ using polar coordinates imposing complex roots,
\begin{equation*}
g_t = 2Q\cos (2 \pi / \gamma) g_{t-1}-Q^2 g_{t-2}+\eta_{gt}.
\end{equation*}

Hereby, $Q$ determines the amplitude and $\gamma$ the frequency of the cycle. The parameterization has the convenient property that available information on the duration and intensity of business cycles can be introduced with relative ease. Incorporating such information using normally distributed priors is complicated, since autoregressive coefficients are more difficult to interpret in terms of the intensity and frequency of the time series. Moreover, allegedly weakly informative Gaussian priors could introduce information on functions of the parameters, potentially placing too much prior weight on dynamics that do not fit observed behavior of output at business cycle frequencies \citep[for a more detailed discussion, see][]{planas2008bayesian}.

Turning to the state equation errors, we assume the elements of $\bm{\eta}_t$ in \autoref{eq: stateeq} to be blockwise orthogonal and achieve this by employing a flexible factor stochastic volatility structure \citep[see, e.g.,][]{aguilar2000bayesian},
\begin{align}
\bm{\eta}_{yt} &= \bm{\Lambda}_y \bm{z}_{yt} + \bm{\varepsilon}_{y t}, \quad \bm{z}_{yt} \sim \mathcal{N}(\bm{0},\bm{\Upsilon}_{yt}), \quad \bm{\varepsilon}_{y t} \sim \mathcal{N}(\bm{0},\bm{\Omega}_{yt}),\label{eq:etayt}\\
\bm{\eta}_{\pi t} &= \bm{\Lambda}_\pi \bm{z}_{\pi t} + \bm{\varepsilon}_{\pi t}, \quad \bm{z}_{\pi t} \sim \mathcal{N}(\bm{0},\bm{\Upsilon}_{\pi t}), \quad \bm{\varepsilon}_{\pi t} \sim \mathcal{N}(\bm{0},\bm{\Omega}_{\pi t}),\label{eq:etapit}\\
\eta_{gt} &\sim \mathcal{N}(0,\text{e}^{\omega_{gt}}).
\end{align}
Here, $\bm{z}_{kt}$ denotes a $q$-dimensional vector of normally distributed latent factors (for $k \in \{y, \pi \}$) with diagonal  $q\times q$-dimensional variance-covariance matrix $\bm{\Upsilon}_{kt} = \text{diag}(\text{e}^{\upsilon_{k1t}},\hdots,\text{e}^{\upsilon_{kqt}})$, and $\bm{\Lambda}_k$ is an $N\times q$ matrix of factor loadings. The idiosyncratic error term $\bm{\varepsilon}_{k t}$ is also Gaussian, with zero mean and  diagonal $N\times N$ variance-covariance matrix $\bm{\Omega}_{k t} = \text{diag}(\text{e}^{\omega_{k1t}},\hdots,\text{e}^{\omega_{kNt}})$. It is noteworthy that any common movements in the innovations determining potential output and trend inflation is purely driven by the latent factors. The presence of $\bm \varepsilon_{kt}$ implies that our model is flexible to allow for country-specific deviations. 

The factor model on the shocks to the states is a parsimonious way of modeling a time-varying variance-covariance matrix since  $q \ll N$. To see this, consider
\begin{align}
\bm{\Upsilon}_t = \begin{bmatrix}
\bm{\Upsilon}_{y t} & \bm{0} & \bm{0}\\
\bm{0} & \bm{\Upsilon}_{\pi t} & \bm{0}\\
\bm{0} & \bm{0} & 0
\end{bmatrix}, \quad
\bm{\Omega}_t = \begin{bmatrix}
\bm{\Omega}_{y t} & \bm{0} & \bm{0}\\
\bm{0} & \bm{\Omega}_{\pi t} & \bm{0}\\
\bm{0} & \bm{0} & \text{e}^{\omega_{gt}}
\end{bmatrix}, \quad
\bm{\Lambda} = \begin{bmatrix} 
\bm{\Lambda}_{y} & \bm{0} & \bm{0}\\
\bm{0} & \bm{\Lambda}_{\pi} & \bm{0}\\
\bm{0} & \bm{0} & 0
\end{bmatrix}.\label{eq:decomposition}
\end{align}
Using \autoref{eq:decomposition}, the $M \times M$ time-varying variance covariance matrix (with $M=2N+1$) of $\bm{\eta}_t$ in \autoref{eq: stateeq} is given by 
\begin{equation*}
\bm{\Sigma}_t = \bm{\Lambda} \bm{\Upsilon}_t \bm{\Lambda}' + \bm{\Omega}_t.
\end{equation*}
Consequently, $\bm{\Sigma}_t$ is block-diagonal, allowing for non-zero covariances of the trend components for output and inflation across countries, respectively, while we impose orthogonality on the trend and cycle components $\bm{\tau}_{yt}$, $\bm{\tau}_{\pi t}$ and $g_t$ across variable types \citep[similar to the assumption introduced by][in the context of single-country output gap estimation]{stock1999forecasting,stock2007has}. For convenience, we define $\bm{z}_t = (\bm{z}_{yt}',\bm{z}_{\pi t}',0)$.

The law of motion imposed on the variances in \autoref{eq: measurement-err} and \autoref{eq: stateeq} remains to be specified. Here we assume that the logarithmic  volatilities in  $\bm{\Upsilon}_t$, $\bm{\Omega}_t$, and  $h_{kit}$ follow independent AR(1) processes.  Specifically, the log-volatility in the measurement equations is given by
\begin{equation}
h_{kit} = \mu_{hki}+\varrho_{hki} (h_{kit-1}-\mu_{hki})+\nu_{kit}, \quad \nu_{kit}\sim \mathcal{N}(0,\vartheta_{hki}).\label{eq: state_hk}
\end{equation}

Using $l=1,\hdots,2q$ and $j=1,\hdots,M$ to indicate the corresponding diagonal element in $\bm{\Upsilon}_t$ and $\bm{\Omega}_t$,  the log of the variances in the state equation evolve according to:
\begin{align}
\upsilon_{lt} 	&= \mu_{\upsilon l} + \varrho_{\upsilon l} (\upsilon_{lt-1}-\mu_{\upsilon l})+\nu_{lt}, 	&&\nu_{lt} \sim \mathcal{N}(0,\vartheta_{\upsilon l}),\\
\omega_{jt} 	&= \mu_{\omega j}+\varrho_{\omega j} (\omega_{jt-1}-\mu_{\omega j})+\nu_{jt}, 				&&\nu_{jt} \sim \mathcal{N}(0,\vartheta_{\omega j}).\label{eq: state_omega}
\end{align}
To simplify notation in the following, we let $\bullet$ denote a placeholder for the various possible combinations of indices. The autoregressive parameters are given by $\varrho_\bullet$, while the means of the log-volatility processes are denoted by $\mu_\bullet$. Finally, the state innovation variances are given by $\vartheta_\bullet$. It is worth noting that if a given $\vartheta_\bullet$ equals zero, the corresponding variance in the measurement or state equation is constant.  Selecting whether equations exhibit time variation in the error variances can thus be carried out efficiently using the techniques stipulated in  \cite{fruhwirth2010stochastic}.


\subsection{Bayesian inference}	\label{sec:Bayesinf}
The model outlined above is quite flexible but also heavily parameterized. This calls for regularization via Bayesian shrinkage priors. We start by outlining a general strategy to shrink our proposed factor model towards a simpler specification when it comes to deciding which components should feature conditional heteroscedasticity. The prior setup on the remaining free coefficients of the model completes this subsection.

In the following we describe how to flexibly select which equations should feature time variation in the variances by shrinking innovation variances in the stochastic volatility specifications to zero. Shrinkage to homoscedasticity in the observation equation is achieved in a similar manner. We start by substituting \autoref{eq:etayt} and \autoref{eq:etapit} in \autoref{eq: stateeq} and then proceed by squaring and taking logs of the $r$th equation ($r = 1,\hdots, M$) to obtain  the non-centered parameterization of the state space model \citep{fruhwirth2010stochastic, kastner2014ancillarity},
\begin{align}
\tilde{\varepsilon}_{rt}& = \mu_{\omega r} + \sqrt{\vartheta_{\omega r}} \tilde{\omega}_{rt} + v_{rt},\quad v_{rt} \sim \ln \chi(1)\\\label{eq: non-centeredSV}
\tilde{\omega}_{rt} &= {\varrho}_{\omega r} \tilde{\omega}_{rt-1} + w_{rt},\quad  w_{rt} \sim \mathcal{N}(0,1),\\
\tilde{\omega}_{rt} &= \frac{\omega_{rt}-\mu_{\omega r}}{\sqrt{\vartheta_{\omega r}}},
\end{align}
with $\tilde{\varepsilon}_{rt}=\ln(\bm{f}_{rt}-\bm{\Phi}_{r \bullet} \bm{F}_t - [\bm{\Lambda}\bm{z}_{t}]_{r\bullet})^2$, $\bm{\Phi}_{r \bullet}$ selecting the $r$th row of the matrix $\bm{\Phi}$, and $[\bm{\Lambda}\bm{z}_{t}]_{r\bullet}$ indicating the $r$th row of $\bm{\Lambda}\bm{z}_{t}$. \Cref{eq: non-centeredSV} implies that the process variance $\vartheta_{\omega r}$ as well as the unconditional mean $\mu_{\omega r}$ is moved from the stochastic volatility state equation into \autoref{eq: stateeq}. Conditional on the full history of the normalized log-volatilities and employing a mixture approximation to render \autoref{eq: non-centeredSV} conditionally Gaussian \citep{kim1998stochastic}, the process variances and parameters can be obtained by estimating an otherwise standard Bayesian linear regression model.

This implies that standard shrinkage priors can be specified on $\sqrt{\theta_{\omega r}}$. We adopt a flexible global-local shrinkage prior proposed in \cite{griffin2010inference} that was recently adopted for state space models in \cite{bitto2016achieving}. Here,
\begin{align*}
\vartheta_{\omega r} \sim \mathcal{G}\left(\frac{1}{2}, \frac{1}{2 B_{\omega r}}\right) \quad \Leftrightarrow \quad \sqrt{\vartheta_{\omega r}} \sim \mathcal{N}(0, B_{\omega r}),
\end{align*}
with a local shrinkage hyperparameter
\begin{equation*}
B_{\omega r} \sim \mathcal{G}(\kappa_\omega, \kappa_\omega \xi_\omega /2), \quad \xi_\omega \sim \mathcal{G}(c_0, c_1).
\end{equation*}
$\xi_\omega$ is the global shrinkage parameter that pushes $\sqrt{\bm{\vartheta}_{\omega}} = (\sqrt{\vartheta_{\omega1}}, \hdots, \sqrt{\vartheta_{\omega M}})'$ towards the prior mean. The hyperparameters $\kappa_\omega$ and $c_0, c_1$ are specified by the researcher. Intuitively, the global shrinkage parameter exerts shrinkage towards the zero vector, while $B_{\omega r}$ serves to pull elements of $\sqrt{\bm{\vartheta}_\omega}$ away from zero when $\xi_\omega$ is large (i.e. heavy global shrinkage is introduced) if supported by likelihood information. We choose an analogous setup for the innovations driving the variances of the latent factors in $\bm{z}_t$,
\begin{equation*}
\sqrt{\vartheta_{\upsilon l}} \sim \mathcal{N}(0, B_{\upsilon l}), \quad B_{\upsilon l} \sim \mathcal{G}(\kappa_\upsilon, \kappa_\upsilon \xi_\upsilon/2), \quad \xi_\upsilon \sim \mathcal{G}(d_0, d_1).
\end{equation*}

The same prior choice is also employed for the process innovation variances in the log volatility equations for the measurement errors,
\begin{equation*}
\sqrt{\vartheta_{hki}} \sim \mathcal{N}(0, B_{hki}), \quad B_{hki} \sim \mathcal{G}(\kappa_h, \kappa_h \xi_h/2), \quad \xi_h \sim \mathcal{G}(e_0, e_1).
\end{equation*}
Notice that the common parameter $\xi_h$ pools information on error variances in the log-volatilities across all output and inflation equations, effectively introducing global shrinkage across variable types. \citet{bitto2016achieving} refer to this prior as a double Gamma prior. Consistent with the literature, we set $\kappa_\omega=\kappa_\upsilon=\kappa_h=0.1$ and $c_0=c_1=d_0=d_1=e_0=e_1=0.01$. This choice introduces heavy shrinkage on all process variances while maintaining heavy tails in the underlying marginal prior, and completes the specifcation to stochastically select the innovation variances that potentially result in heteroscedastic errors in both the observation and state equations.

Priors on the factor loadings in \autoref{eq: obs_output}, \autoref{eq: obs_pi} and \autoref{eq:decomposition}, the free autoregressive coefficient in \autoref{eq: stateeq}, and the stochastic volatility parameters remain to be specified. Following \cite{planas2008bayesian}, we specify a Beta distributed prior on the amplitude $Q$ of the business cycle,
\begin{equation}
Q \sim \mathcal{B}(a_Q, b_Q),\label{priorQ}
\end{equation}
with $a_Q = 5.82$ and $b_Q = 2.45$ denoting hyperparameters chosen specifically for euro area business cycles. For the frequency $\gamma$ we also adopt a Beta prior with
\begin{equation}
\frac{\gamma-\gamma_L}{\gamma_H-\gamma_L} \sim \mathcal{B}(a_\gamma, b_\gamma).\label{priortau}
\end{equation}
This prior restricts the support of $\gamma$ by specifying a minimum wave length $\gamma_L$, which is set equal to two, and a maximum length $\gamma_H$ set equal to $T$. The parameters $a_\gamma = 2.96$ and $b_\gamma = 10.7$ are fixed hyperparameters again set specifically according to prior research on business cycles in the euro area. These choices are weakly informative and imply a periodicity of around eight years and a contraction factor of $0.8$ \citep{gerlach1999output,planas2008bayesian,jarocinski2018inflation}.

For the remaining parameters of Eqs. (\ref{eq: state_hk}) to (\ref{eq: state_omega}) we follow \cite{kastner2014ancillarity} and use a weakly informative Gaussian prior on the unconditional means, $\mu_{\bullet} \sim \mathcal{N}(0, 10^2)$ as well as a Beta prior on the persistence parameter $\varrho_{\bullet} \sim \mathcal{B}(25, 5)$. On the factor loadings $\alpha_{ki}$ and $\beta_{ki}$ that reflect the sensitivity of country-specific output and inflation measures to the cycle components, we use a sequence of independent Gaussian priors with $\alpha_{ki} \sim \mathcal{N}(0,1)$ and $\beta_{ki} \sim \mathcal{N}(0,1)$. For the factor loadings in $\bm{\Lambda}_k$ governing the covariance structure for the trend components across countries, with $\lambda_{k\bullet}$ indicating the elements, we use tight independent Gaussian priors $\lambda_{k\bullet} \sim \mathcal{N}(0,0.1)$. Finally, we specify the priors on the initial state $\bm{f}_0$ and the log-volatilities to be fairly uninformative with each element being normally distributed with zero mean and variance $10^2$.

Notice that some parts of the parameter space of the model specified above are not econometrically identified. In the measurement equation, to identify the scale and sign of the output gap, we normalize the loading for the first country using the restriction $\alpha_1 = 1$. Moreover, we restrict the factor loadings matrices $\bm{\Lambda}_k$ following \citet{aguilar2000bayesian} by setting the respective upper $q\times q$ blocks equal to lower triangular matrices with ones on the main diagonals.

These priors are then combined with the likelihood to obtain the posterior distribution. Since the joint posterior is intractable, we employ a Markov chain Monte Carlo (MCMC) algorithm detailed in \cref{sec:MCMC algorithm}. This algorithm samples all coefficients and latent quantities from their full conditional posterior distributions to obtain, after a potentially large number of iterations, valid draws from the joint posterior density. The algorithm is repeated $50,000$ times with the first $25,000$ draws discarded as burn-in. Convergence and mixing of most model parameters appear to be satisfactory. However, we find a substantial degree of autocorrelation for the factor loadings in selected countries. To assess the sensitivity of our findings, we thus re-estimated the model a moderate number of times based on different initial values. The corresponding findings appear to be remarkably robust.

\section{In-sample features of the model}\label{sec: empirics}

\subsection{Data overview}\label{sec: data_overview}
For the empirical application, we use quarterly data for economic output, measured in terms of real gross domestic product (RGDP, seasonally adjusted), and the harmonized index of consumer prices (HICP, in year-on-year growth rates), respectively. To obtain a measure of the output gap in percent, we transform the output variable by applying the transformation $400\log(RGDP)$. We choose $q=1$ latent factors for both the potential output for all countries and trend inflation measures to capture the covariances between the country-specific quantities.

In terms of time and country coverage, our sample runs from 1997:Q2 to 2018:Q4. Thus, it includes the period surrounding the introduction of the euro, the global financial crisis (GFC) in 2008/2009, and the more recent crisis of the euro area peripheral countries. To achieve consistent data coverage, we include Austria (AT), Belgium (BE), Finland (FI), France (FR), Germany (DE), Greece (GR), Italy (IT), Netherlands (NL), Portugal (PT) and Spain (ES), resulting in $N = 10$.

\subsection{Euro area output gap estimates}\label{sec: output_gap}
\begin{figure}
\begin{center}
\includegraphics[width=0.8\textwidth]{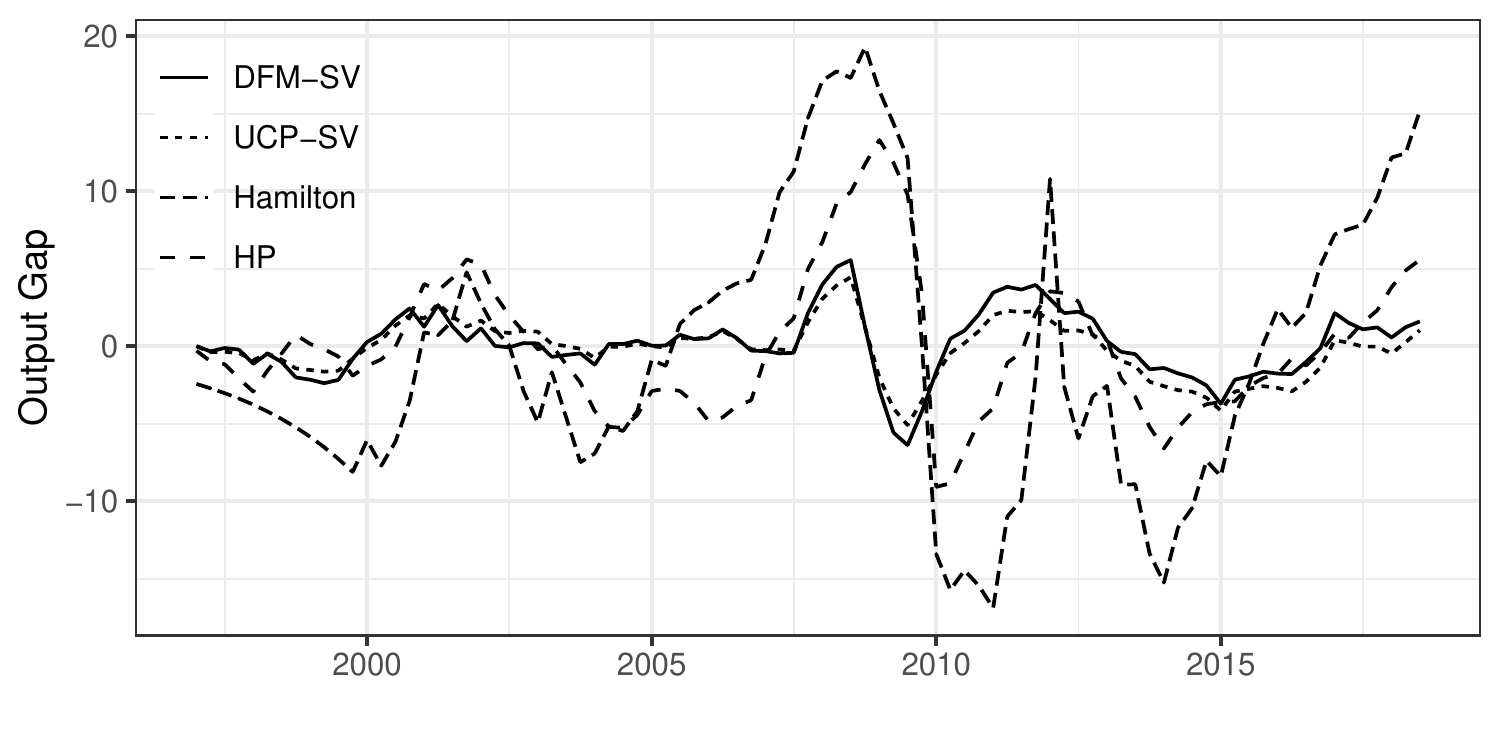}
\end{center}
\caption{Competing approaches to measuring the euro area output gap.}\vspace*{0.25em}
\footnotesize\textit{Notes}: Dynamic factor model with stochastic volatility (DFM-SV) is the approach set forth in this paper exploiting cross-sectional information; unobserved components model with stochastic volatility (UCP-SV) refers to a standard specification based on aggregate euro area data. Hamilton denotes the approach set forth in \citet{hamilton2017you}, while the remaining specification is the Hodrick-Prescott filter \citep[HP,][]{hodrick1997postwar}. Lines indicate the respective estimated posterior median.
\label{fig:competitors}
\end{figure}

In this subsection we present some key in-sample results of our proposed model. We start by comparing the estimated output gap with other competing measures, which are depicted in  \autoref{fig:competitors}. The black line in  \autoref{fig:competitors} shows the posterior median of the euro are output gap using the model specification sketched above (DFM-SV).  To assess whether using cross-sectional information on prices and output leads to significantly different conclusions, we include a model similar to the one proposed but exclusively relying  on aggregate data for the EA (labeled UCP-SV). This model is closely related to the multivariate unobserved components model proposed in \cite{stella2013state}.  Furthermore, to inspect whether our state evolution specification yields different dynamics in the gap component, we also include two model specifications that replace $g_t$ with a plug-in estimate $\hat{g}_t$. As estimators for $g_t$, we use the approach proposed in \cite{hamilton2017you} (labeled Hamilton) and the well-known Hodrick-Prescott filter (HP, \citealt{hodrick1997postwar}) as a means to dissecting economic output series into a trend and a cyclical component. These gap terms are computed based on aggregated data and then included in the model described in \cref{sec: econometrics}.

\Cref{fig:competitors} indicates that the output gap obtained from our multi-country framework closely tracks the output gap measure obtained by estimating a bivariate unobserved components model based on aggregate data, especially in the beginning of the sample. During the GFC, we observe a slight decoupling in terms of gap estimates between DFM-SV and the UCP-SV. Comparing both output gap measures with estimates arising from a model based on the approach proposed in \cite{hamilton2017you} and a standard HP filter yields  several interesting insights. 

First, the Hamilton gap indicates that in the end of the 1990s, output in the EA has been consistently below potential output until the early 2000s. This pronounced negative gap is not visible for any of the remaining three approaches. Second, the Hamilton and the HP measure indicate a strong positive deviation of output from trend output in the run-up to the GFC with a slightly delayed but sharp drop in the final quarter of 2008. By contrast, our proposed measure already drops in the first half of 2008 while a turning point in the business cycle is visible from mid 2009 onwards. At a first glance, it seems that this earlier drop in the output gap and the more timely rebound in real activity can be traced back to the fact that cross-sectional information is efficiently exploited. However, it is noteworthy that the measure based on the UCP-SV model also tends to react faster compared to Hamilton and HP. Since this model, as opposed to DFM-SV, is not exploiting cross-sectional information explicitly, we conjecture that the more timely reaction might come from modeling real activity and prices jointly. Third, and finally, notice that both measures based on unobserved components models exhibit a significantly smaller volatility and appear to be smoother. This effect is mainly due to our prior setup that  softly introduces smoothness as well as additional information on the length and intensity of the cycle. 

We close this subsection by reporting prior and posterior summary statistics of the amplitude $Q$ and frequency $\gamma$, depicted in Table \ref{tab:ar2par}. The table shows means and standard deviations associated with the prior and posterior of $Q$ and $\gamma$, respectively. This comparison allows us to assess how much information on the shape of the output gap is contained in the likelihood and, in addition, enables a comparison to the results reported in \cite{planas2008bayesian}. Considering the posterior mean and standard deviation of $\gamma$ suggests that the average length of the cycle is about 6.5 years. For data spanning from the 1980s to the early 2000s, \cite{planas2008bayesian} report significantly longer cycles. Since our sample period covers the GFC as well as the EA periphery crisis, this finding is not surprising since both shocks lead to abrupt downward movements in the business cycle.  Comparing the prior and posterior dispersion indicates that the information contained in the prior is not reducing estimation uncertainty significantly. 

 Next, we discuss the intensity of business cycle movements by considering the amplitude $Q$. Compared to previous studies, our estimate appears to be slightly lower. Since \cite{planas2008bayesian} rely on aggregate data, the lower value of $Q$ can be explained by the fact that our aggregate gap measure strikes a balance between capturing the higher business cycle variance of EA peripheral countries such as Greece and Spain while capturing information on more stable business cycles found in, e.g., Germany and Austria. Note that the prior and posterior mean are close to each other but the prior and posterior standard deviations differ strongly.  This highlights that the introduction of prior information helps in reducing posterior uncertainty.

\begin{table*}
\caption{Prior and posterior moments of the AR(2)-process parameters.}\label{tab:ar2par}\vspace*{-1.8em}
\begin{center}
\begin{threeparttable}
\small
\begin{tabular*}{0.5\textwidth}{c @{\extracolsep{\fill}} lrrrr}
  \toprule
& \multicolumn{2}{c}{Prior} & \multicolumn{2}{c}{Posterior}\\
\cmidrule(l{0pt}r{3pt}){2-3}\cmidrule(l{3pt}r{0pt}){4-5}
 & Mean & SD & Mean & SD \\ 
  \midrule
  Frequency $\gamma$ & 20.40 & 9.15 & 26.07 & 9.63 \\ 
  Amplitude $Q$ & 0.70 & 0.15 & 0.68 & 0.07 \\ 
   \bottomrule
\end{tabular*}
\begin{tablenotes}[para,flushleft]
\scriptsize{\textit{Notes}: SD -- Standard deviation. Summary statistics refer to the prior moments in Eqs. (\ref{priorQ}) to (\ref{priortau}). Posterior indicates the measures obtained from the posterior draws.}
\end{tablenotes}
\end{threeparttable}
\end{center}
\end{table*}

\subsection{The role of stochastic volatility in modeling the output gap}\label{sec: SVrole}
\begin{figure}
\begin{subfigure}{.62\textwidth}
\begin{center}
\includegraphics[width=\textwidth]{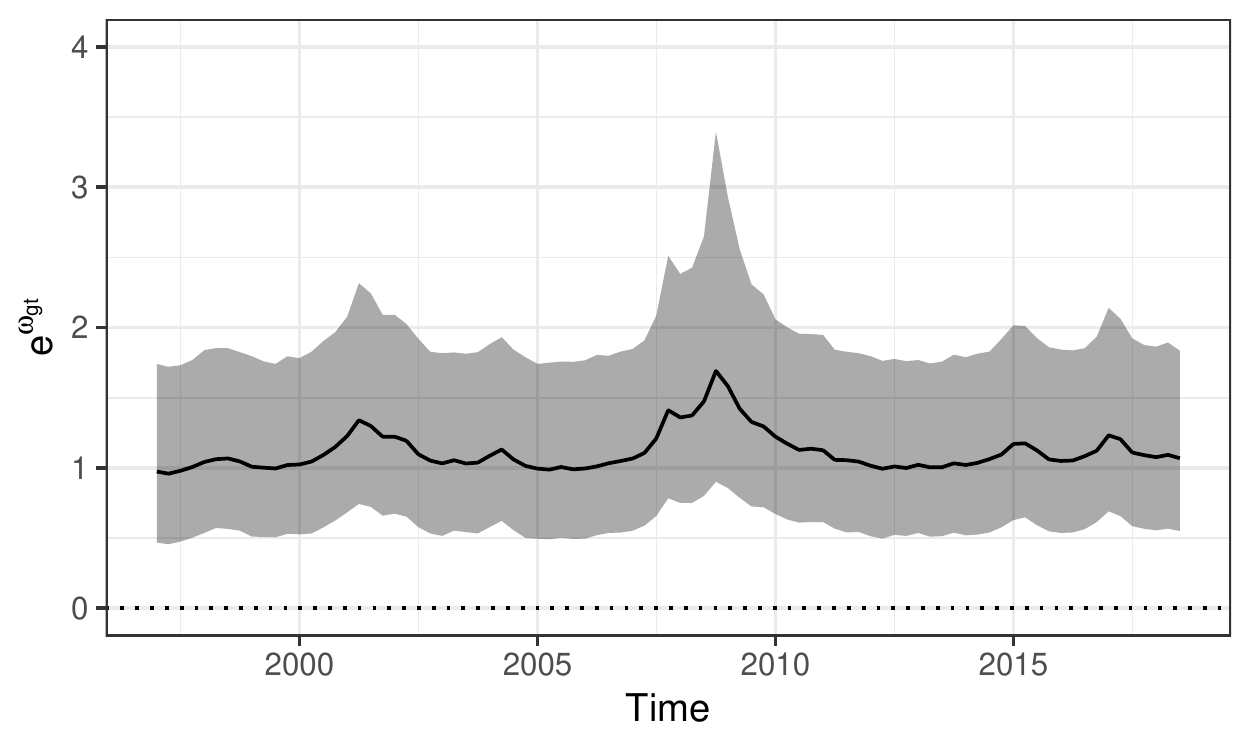}
\end{center}
\end{subfigure}
\begin{subfigure}{.37\textwidth}
\begin{center}
\includegraphics[width=\textwidth]{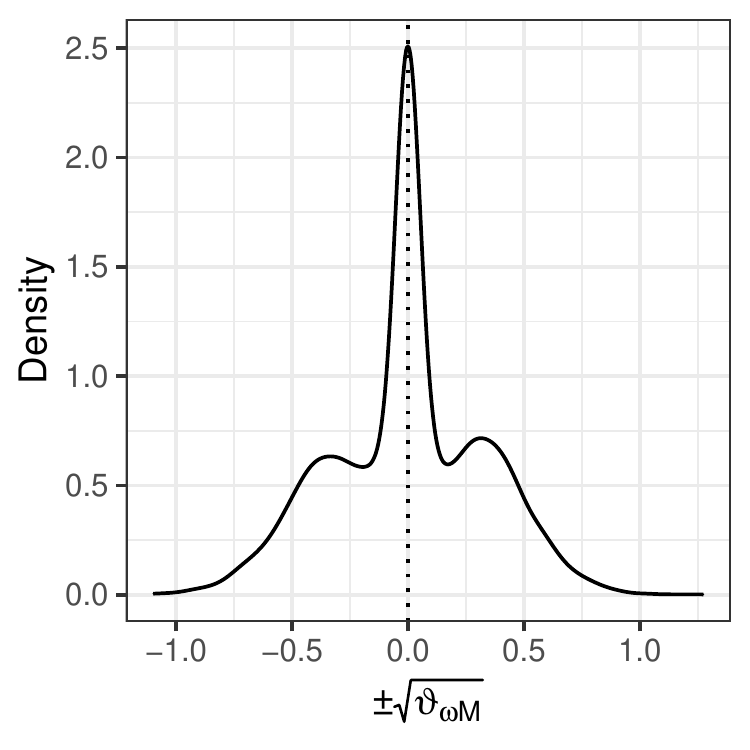}
\end{center}
\end{subfigure}
\caption{Stochastic volatility of the euro area output gap.}\vspace*{0.25em}
\footnotesize \textit{Notes}: The left panel depicts the posterior variance of the output gap component over time. The solid line indicates the estimated posterior median, with grey shaded areas covering the area between the $16$th and $84$th percentile. The right panel shows a kernel density estimate of the posterior distribution of the signed square root of the innovation variance to the stochastic volatility process of the output gap. The dotted line marks zero.
\label{fig:gapsv}
\end{figure}

In the next step, we ask whether the volatility of the shocks driving the area-wide output gap is time-varying. To this end,  the left panel in  \autoref{fig:gapsv} displays the posterior median of the stochastic volatility component of the euro area output gap of our proposed multi-country model DFM-SV along with the lower 16th and upper 84th percentile of the credible interval (grey shaded area). Considering the posterior quantiles in \autoref{fig:gapsv} provides some limited evidence in favor of heteroscedasticity. We observe slight increases during the burst of the dot-com bubble as well as during the period of the GFC. 

One way of assessing the likelihood that heteroscedasticity in the business cycle shocks is present is to consider the posterior distribution of the square root of $\vartheta_{\omega M}$ up to a random sign switch. In case of homoscedasticity, the corresponding marginal posterior would be unimodal and centered on zero. Consideration of the right panel of \autoref{fig:gapsv} corroborates the discussion above, namely that evidence for heteroscedasticity is, at best, limited. While the marginal posterior is clearly not unimodal, most posterior mass is located around zero.

\begin{figure}
\includegraphics[width=\textwidth]{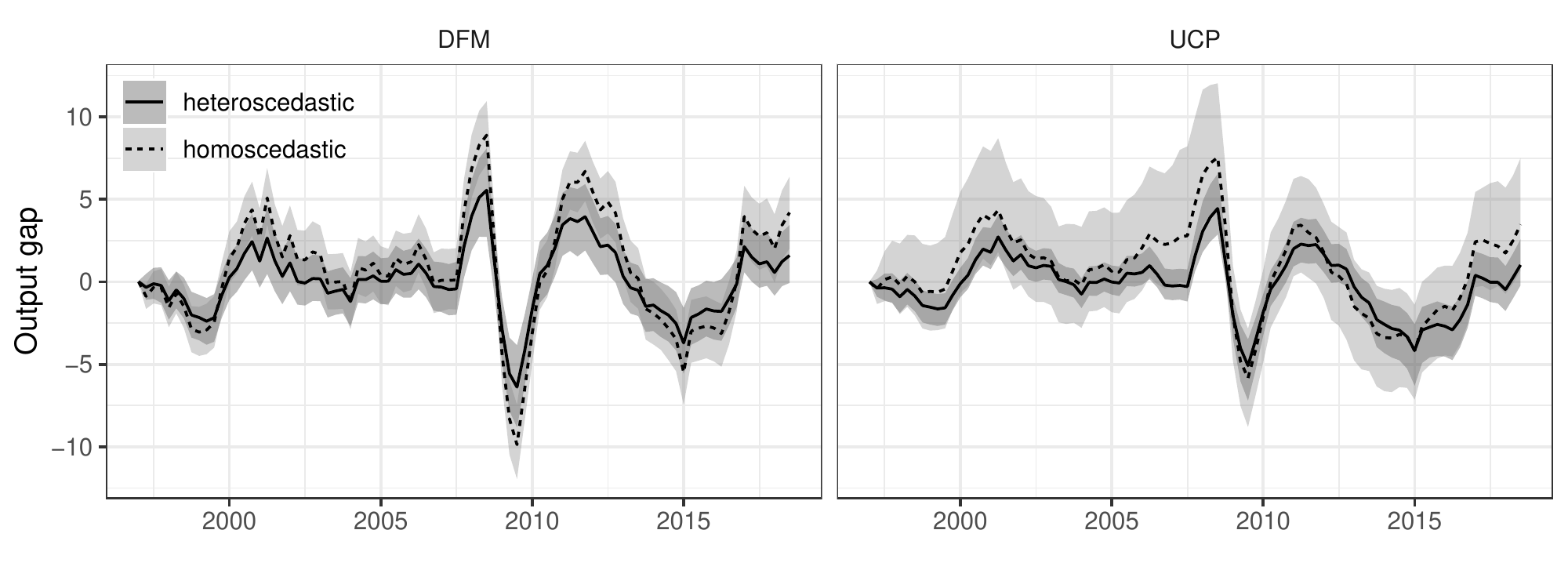}
\caption{Dynamic factor and unobserved component models with and without stochastic volatility.}\vspace*{0.25em}
\footnotesize \textit{Notes}: Dynamic factor model (DFM) is the approach set forth in this paper exploiting information across euro area countries. Unobserved components model (UCP) refers to a standard specification based on aggregate euro area data. Solid and dashed lines indicate the estimated posterior median, with grey shaded areas covering the area between the $16$th and $84$th percentile. 
\label{fig:svvsnosv}
\end{figure}

To assess how the presence of stochastic volatility in the unobserved components impacts the estimate of the output gap, \Cref{fig:svvsnosv} shows the posterior median of the output gap under our baseline specification (in solid black) alongside the 16th and 84th percentiles (dark shaded area) for the DFM and the UCP model. The dashed black line represents the posterior median of the output gap obtained by estimating the model without stochastic volatility for all latent components, with the light gray area denoting the 16th and 84th percentiles. One key finding of this figure is that for the DFM, switching off SV yields a similar measure of the output gap that is quite close to the one obtained under the DFM with SV. The main differences concern the magnitude and variability of the gap measure. Put differently, comparing the posterior median across the two specifications points towards more pronounced movements in $g_t$ obtained from the model without stochastic volatility. This finding is closely related to the critique raised by \cite{sims2001comment} and \cite{stock2001evolving} in response to the work of \cite{cogley2001evolving}, who estimate a time-varying parameter model without stochastic volatility. Ignoring stochastic volatility, within the framework of a time-varying parameter model, is expected to exaggerate movements in the regression coefficients and potentially creating transient variations in filtered estimates.

\subsection{Dissecting euro area business cycle movements}\label{sec: dissecting}
\begin{figure}[ht!]
\begin{center}
\includegraphics[width=\textwidth]{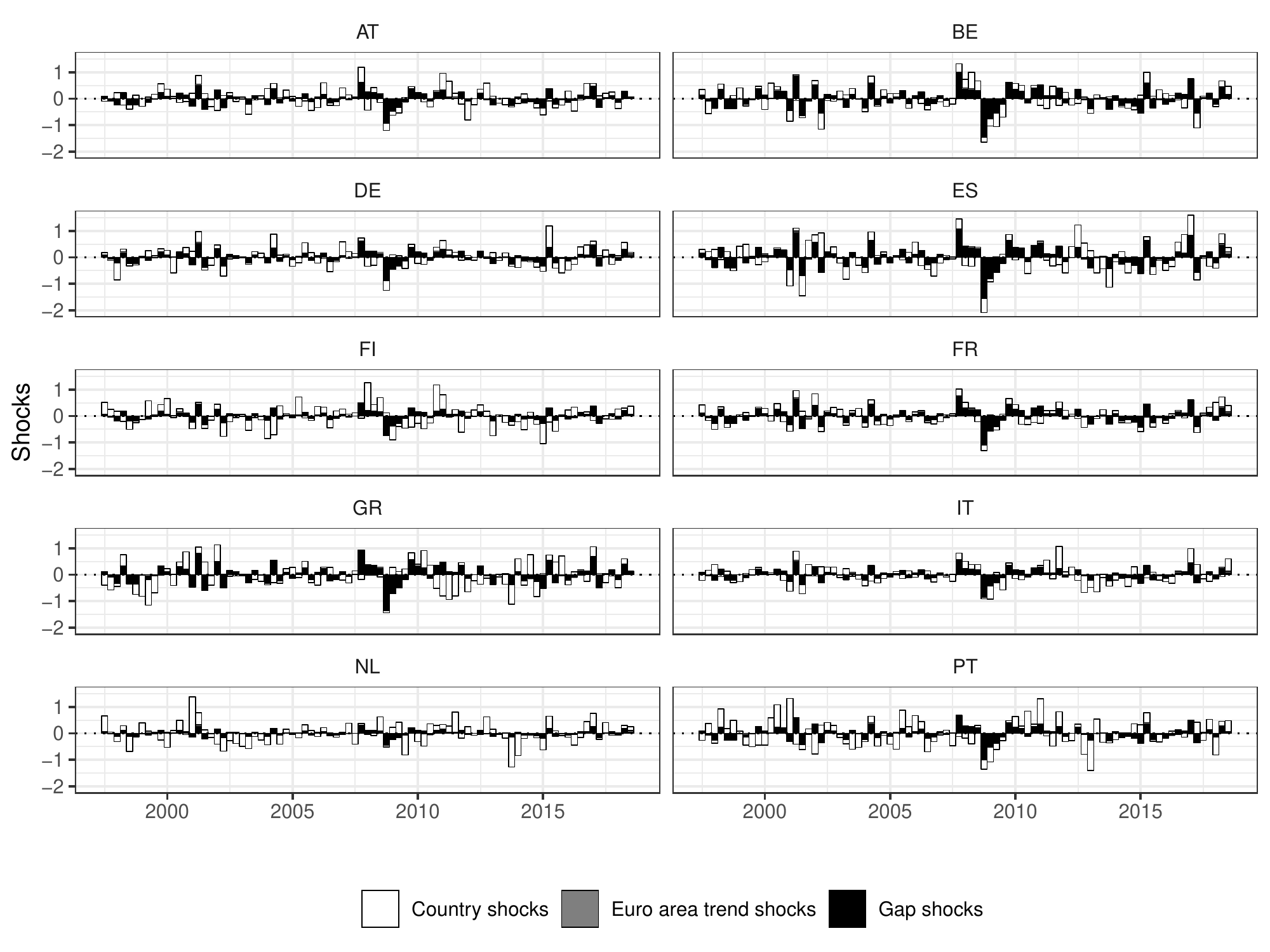}
\end{center}
\vspace*{-1em}\caption{Decomposing shocks shaping inflation across countries.}\vspace*{0.25em}
\footnotesize \textit{Notes}: Shocks refer to the posterior median of the estimated shocks of the fitted model. \textit{Country shocks} are shocks specific to all countries and thus include the idiosyncratic component of the factor stochastic volatility specification and the measurement errors. The remaining quantities arise from joint euro area dynamics; \textit{Euro area trend shocks} refer to shocks identified based on common euro area factors underlying country-specific potential output, while \textit{Gap shocks} for country $i$ arise solely from the gap component. The dotted line marks zero.
\label{fig:shocksDp}
\end{figure}

In the following, we provide information on the quantitative contributions of shocks to trend, cyclical and idiosyncratic components to the observed series of inflation over time. Here we use an approach similar to a standard historical time series decomposition. Notice that the non-stationary nature of the trend components in \autoref{eq: stateeq} implies that shocks to these quantities are persistent and do not peter out.  In fact, instead of becoming less important over time, the relative importance of shocks to the trend components increases by construction. As a consequence, we focus on the contributions of the shocks at each point in time. Combining Eqs. (\ref{eq: obs_pi}),  (\ref{eq: stateeq}) and (\ref{eq:etapit}), we can decompose inflation across countries in terms of their shocks and lagged states:
\begin{equation*}
{\pi}_{it} =\tau_{\pi it-1}+ \beta_i \phi_1 g_{t-1} + \beta_i \phi_ 2 g_{t-2}+ [\bm{\Lambda}_\pi\bm{z}_{\pi t}]_{i\bullet} + [\bm{\varepsilon}_{\pi t}]_i + \beta_i \eta_{gt} + \epsilon_{\pi it}\label{eq:shockdecompDp}.
\end{equation*} 
The decomposition yields three individual shocks of interest, with $[\bm{\Lambda}_\pi \bm{z}_{\pi t}]_{i\bullet}$ and $\beta_i \eta_{gt}$ reflecting joint area-wide dynamics, while $[\bm{\varepsilon}_{\pi t}]_i$ and $\epsilon_{\pi it}$ capture idiosyncratic shocks. In particular, $[\bm{\Lambda}_\pi \bm{z}_{\pi t}]_{i\bullet}$ arises from the factor stochastic volatility structure and indicates common euro area trend component shocks (subsequently labeled \textit{Euro area trend shocks}). The contribution of the gap component is given by $\beta_i \eta_{gt}$ (indicated as \textit{Gap shocks} in the following). The quantity $[\bm{\varepsilon}_{\pi t}]_i$ is a country-specific shock to the trend component, while $\epsilon_{\pi it}$ is the idiosyncratic measurement error (labeled \textit{Country shocks} and considered jointly in what follows). To ease visualization, \autoref{fig:shocksDp} shows the posterior median of period-specific shocks exclusively based on $\tilde{\pi}_{it} = \pi_{it}  - \tau_{\pi it-1} - \beta_i \phi_1 g_{t-1} - \beta_i \phi_2 g_{t-2}$.

Figure \ref{fig:shocksDp} reveals a set of interesting results for the shock decomposition of inflation across countries. First, the most striking observation is that \textit{Euro area trend shocks} do not play a role in driving observed inflation series. This finding results from an almost diagonal variance-covariance structure between country-specific trends of inflation, with most covariances rather close to zero. In terms of the modeling setup, this implies that one may safely impose orthogonality on the errors for the trend inflation state equations. 

Second, we find substantial evidence for the existence of a Phillips curve relationship across the EA countries given by \textit{Gap shocks}. Notice that the sensitivity of country-specific inflation series to area-wide output gap shocks is governed by the factor loadings $\beta_i$. Here, we find that the slope of the Phillips curve exhibits heterogeneity, with the Netherlands and Finland providing examples of less sensitive countries. By contrast, the area-wide output gap shocks appear to be particularly important for the dynamic evolution of inflation in Belgium, Spain and Greece. This result implies that almost all comovements in inflation across countries arises from the joint gap component rather than shocks to country-specific trend inflation. 

Finally, we assess the importance of country-level shocks. Recall that these shocks depict both shocks to idiosyncratic trends, but also the measurement errors. It is worth mentioning that measurement errors play only a minor role in shaping the observed inflation series over the cross-section, and the contributions labeled \textit{Country shocks} mainly feature shocks to the trend components. The highest importance of such country-level shocks is apparent for the cases of Greece, Italy and Portugal in the five year period after 2010, while inflation in the Netherlands appears to be shaped to a large extent by idiosyncratic shocks throughout the observed period.

\subsection{Responses of output and inflation to business cycle shocks} \label{sec: shocks}
This subsection aims at studying the dynamic effect of business cycle shocks to inflation across the euro area. Such a common shock is of interest for policy makers in order to assess the sensitivity of their respective countries to common adverse movements in an area-wide business cycle. In our framework, a business cycle shock is defined as an unexpected decrease in $\eta_{gt}$ by one standard deviation. This yields dynamic reactions of $g_{t+h}~(h=1,\dots,H)$ that are then transformed into dynamic reactions of $y_{it+h}$ and $\pi_{it+h}$ by using the factor loadings $\alpha_i$ and $\beta_i$. These impulse response functions (IRFs) thus provide not only information on the specific time profile of the output gap reactions but also on the sensitivity of a given country and variable to such changes.

\begin{figure}[t]
\begin{center}
\includegraphics[width=0.8\textwidth]{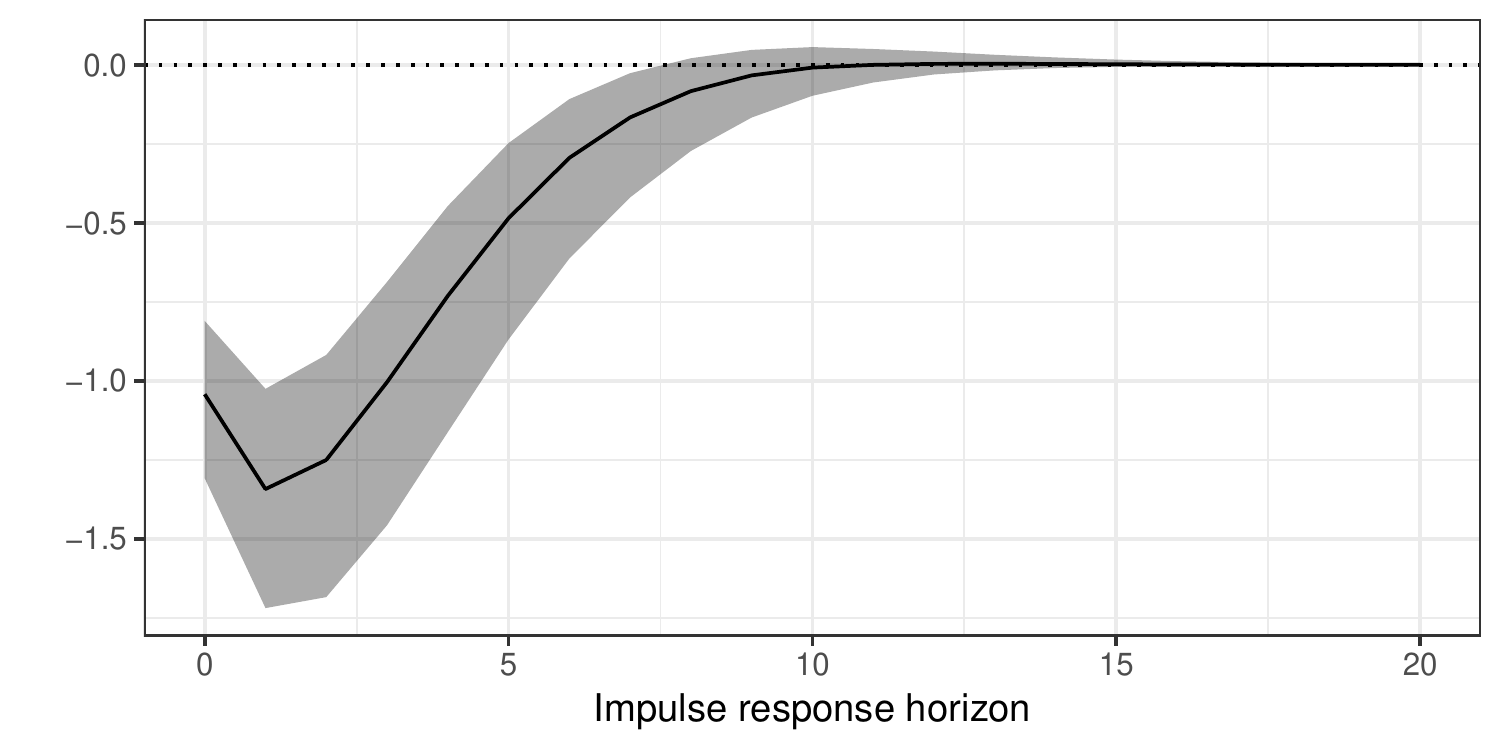}
\end{center}
\vspace*{-1em}\caption{Impulse response function of a negative one standard deviation shock to the output gap.}\vspace*{0.25em}
\footnotesize \textit{Notes}: The solid black line depicts the median response alongside the $16$th and $84$th percentiles shaded in grey. The dotted line marks zero.
\label{fig:irf_gap}
\end{figure}

Figure \ref{fig:irf_gap} depicts the posterior distribution of the IRF of the common output gap to a (negative) one standard deviation business cycle shock. The black line in the figure represents the median responses over time along with lower 16th and upper 84th percentiles of the posterior distribution. The figure presents a negative and immediate impact on the common gap component, with a peak decline in the output gap of around 1.5 percentage points. This peak is reached after around three quarters and rapidly dies out afterwards. After around 2.5 years, the effect on the output gap is zero.

It is worth noting that  \autoref{fig:irf_gap} only measures the dynamic impact to the latent gap component. However, policy makers might be particularly interested in how changes in the common cycle impact inflation across countries. Since the dynamics of $\pi_{it}$ are proportional to movements in $g_t$, we report peak effects that are reached after around three quarters (see \autoref{fig:irf_gap}).


\begin{figure}[t]
\begin{center}
\includegraphics[width=0.8\textwidth]{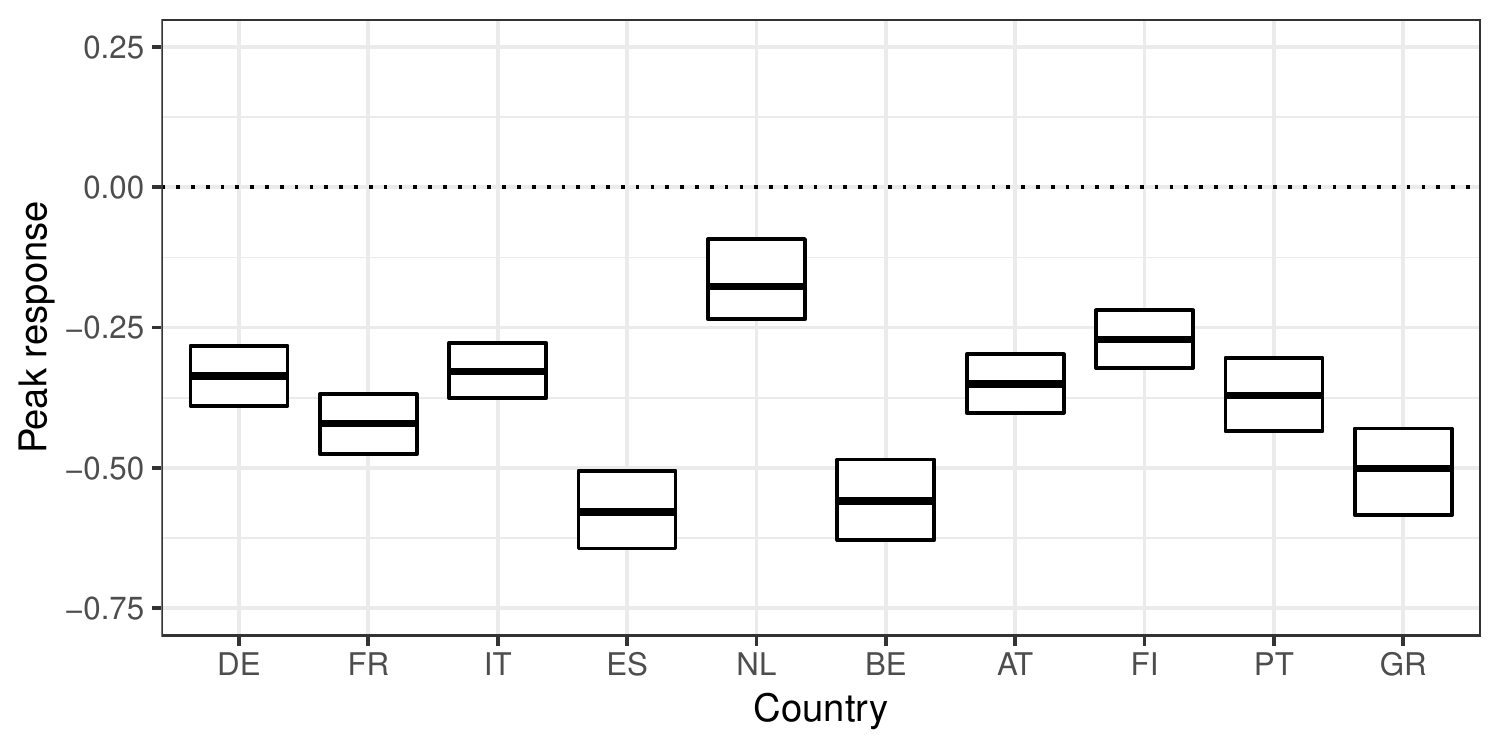}
\end{center}
\vspace*{-1em}\caption{Peak responses of a euro area business cycle shock for inflation across countries.}\vspace*{0.25em}
\footnotesize \textit{Notes}: Boxplots refer to the posterior draws at the peak response. Boxes cover the area between the $16$th and $84$th percentile, with the solid black line depicting the posterior median. The dotted line marks zero.
\label{fig:irf_countries}
\end{figure}

Inspection of the maximum responses of inflation in \autoref{fig:irf_countries} reveals that a common business cycle shock translates into a drop in inflation across all countries under scrutiny. This drop in inflation ranges from about $-0.8$ to approximately $-0.2$ percentage points. These findings are highly significant and provide strong evidence that a Phillips curve relationship exists in the EA. This corroborates recent evidence reported in \citet{onoranteECBWP}, who find a relationship between inflation and real activity based on aggregate EA data. However, notice that there exist considerable differences in peak inflation reactions across countries, which could explain findings in \cite{barigozzi2014euro} and \cite{peersman2004transmission}, who report asymmetric responses of macroeconomic quantities to common monetary policy shocks in the euro area, given that the link between demand-sided policies and inflation differ across EA member states. 

\section{Forecasting evidence}
Up to this point we have focused on in-sample results to illustrate the key features of our proposed modeling approach. However, a successful model that could be useful for policy analysis should also be able to predict well. To investigate the predictive capabilities, this section builds on the literature on inflation forecasting \citep[see][among others]{stock1999forecasting,stock2007has,jarocinski2018inflation,koop2018forecasting} and uses our model to forecast aggregate inflation for the EA and across individual member states up to four quarters ahead.
\subsection{Design of the forecasting exercise and competing models}
To evaluate forecast performance for both the EA and individual countries, we split the sample into an initial estimation period that ranges from 1997:Q2 to 2008:Q3 ($47$ observations) and use the remaining $40$ observations  as a hold-out period. The forecasting design adopted is recursive, implying that after obtaining a set of predictive densities, we increase the length of the initial observation period by one quarter until we reach the end of the hold-out period.

Differences in predictive accuracy are gauged by relying on root mean squared errors (RMSEs) and log predictive scores \citep[LPSs, see][]{geweke2010comparing}. RMSEs are obtained by considering the differences between the posterior median of the predictive distribution and the realized values of $\pi_{it}$ for each model and across the hold-out period. Analogously, LPSs are computed by evaluating the realized values under the predictive density of a given model, summed over the hold-out.

We benchmark the proposed DFM-SV model against a range of competing models that  differ  in several respects. First, we distinguish between models that exploit cross-sectional information (labeled \textit{Multi-country}) versus specifications that utilize only country-specific information (labeled \textit{Single-country}). In the case of aggregate euro area inflation forecasts, we use the abbreviation \textit{EA-level} to indicate that predictions are based on observations of output and inflation that are aggregated from country-level data prior to estimation to yield a measure of EA-level inflation and output. Here, aggregate refers to taking the arithmetic mean over the respective country-specific series.  Second, we consider a range of alternative measures of the output gap to assess differences between treating the output gap as a latent quantity as opposed to using an observed measure. Third, we gauge the accuracy gains from stochastic volatility by also including homoscedastic variants of all competing models.\footnote{Here, homoscedasticity implies that we assume constant error variances in the state and observation equations.}

The model set we consider is comprised of the following benchmarks:
\begin{enumerate}[label=(\roman*),align=left]
\item Unobserved components model (UCP): This specification refers to a modeling approach in the spirit of \citet{stella2013state}. For \textit{Single-country}, this implies that we introduce country-specific gap components and estimate individual country-specifications with orthogonal error terms. This yields a model setup per country that estimates three unobserved components. By comparing the out-of-sample predictive performance of the UCP specifications with forecasts produced by DFM, the inclusion of these specifications serves to assess the merits of considering multi-country information as a means to improving both country-specific and aggregate predictions. In the case of \textit{EA-level}, we aggregate country-level series a priori and estimate the model using three latent factors.
\item Hamilton (\textit{Ham}): These specifications rely on a plugin-estimate $\hat{g}_t$ as the measure of the output gap in the framework proposed in this paper. For \textit{Ham}, we follow recent work by \cite{hamilton2017you} as a means to estimating the gaps. We calculate forecasts for \textit{Single-country} (extracting gaps for each country individually) and \textit{Multi-country} (aggregating a priori and using EA-level information).
\item Hodrick-Prescott (HP): Similar to the strategy employed for for \textit{Ham}, these specifications use the well-known HP filter (\citealt{hodrick1997postwar}) to produce euro area output gap estimates. For the forecast comparison, again both multi-country and single-country specifications for HP are implemented.
\item AR(1): A standard homoscedastic autoregressive process of order one used to forecast aggregate euro area inflation, and country-specific inflation series independently.
\end{enumerate}
In what follows, all models are benchmarked against  the AR(1) model. Here, we consider relative RMSEs and differences in LPSs of all specifications versus the AR(1) model. RMSEs below 100 thus reflect that the respective model outperforms the benchmark in terms of point predictions, while LPSs exceeding zero indicate superior performance for density forecasts vis-\'{a}-vis the AR(1) specification.

\subsection{Aggregate euro area inflation forecasts}
In this subsection we assess whether our model yields competitive forecasts for aggregate data.  Out-of-sample performance for aggregate euro area inflation is evaluated for the one-quarter up to the four-quarter ahead horizon. Table \ref{tab:inflation-ea} reports  relative RMSEs and differences in LPSs, benchmarked to the AR(1) model.  

\begin{table*}[ht!]
\caption{Forecast evaluation for euro area inflation.}\label{tab:inflation-ea}\vspace*{-1.8em}
\begin{center}
\begin{threeparttable}
\scriptsize
\begin{tabular*}{\textwidth}{@{\extracolsep{\fill}} lrrrrrrrrrrrrrr}
  \toprule
& \multicolumn{6}{c}{Multi-country} & \multicolumn{6}{c}{EA-level}\\
 \cmidrule(l{0pt}r{3pt}){2-7}\cmidrule(l{3pt}r{0pt}){8-13}
& \multicolumn{3}{c}{non-SV} & \multicolumn{3}{c}{SV} & \multicolumn{3}{c}{non-SV} & \multicolumn{3}{c}{SV}\\
 \cmidrule(l{0pt}r{3pt}){2-4}\cmidrule(l{3pt}r{3pt}){5-7}\cmidrule(l{3pt}r{3pt}){8-10}\cmidrule(l{3pt}r{0pt}){11-13}
& DFM & Ham & HP & DFM & Ham & HP & UCP & Ham & HP & UCP & Ham & HP \\ 
\midrule
\textbf{LPS} &  &  &  &  &  &  &  &  &  &  &  &  \\ 
  \textit{1-Qt} & $42.9$ & $40.4$ & $38.6$ & $\bm{43.0}$ & $38.7$ & $40.5$ & $-41.1$ & $-38.1$ & $-33.9$ & $-26.1$ & $-29.8$ & $-32.5$ \\ 
  \textit{2-Qt} & $74.0$ & $72.2$ & $70.0$ & $77.9$ & $76.1$ & $\bm{79.4}$ & $-60.5$ & $-57.2$ & $-53.3$ & $-41.2$ & $-60.7$ & $-69.7$ \\ 
  \textit{3-Qt} & $77.9$ & $74.5$ & $71.8$ & $\bm{85.0}$ & $75.6$ & $81.8$ & $-62.0$ & $-66.8$ & $-60.5$ & $-37.3$ & $-61.2$ & $-97.1$ \\ 
  \textit{4-Qt} & $72.0$ & $67.9$ & $67.0$ & $\bm{76.4}$ & $62.5$ & $73.7$ & $-60.3$ & $-67.8$ & $-59.6$ & $-38.7$ & $-58.9$ & $-89.6$ \\ 
  \midrule
  \textbf{RMSE} &  &  &  &  &  &  &  &  &  &  &  &  \\ 
  \textit{1-Qt} & $97.3$ & $102.6$ & $100.4$ & $\bm{97.0}$ & $120.5$ & $115.7$ & $100.4$ & $99.5$ & $101.0$ & $97.6$ & $115.2$ & $113.7$ \\ 
  \textit{2-Qt} & $97.7$ & $102.7$ & $102.2$ & $\bm{96.2}$ & $104.9$ & $105.7$ & $100.3$ & $99.8$ & $101.8$ & $96.4$ & $105.8$ & $107.2$ \\ 
  \textit{3-Qt} & $97.2$ & $102.1$ & $103.3$ & $95.0$ & $\bm{84.7}$ & $99.3$ & $100.1$ & $99.6$ & $101.9$ & $94.5$ & $100.9$ & $104.1$ \\ 
  \textit{4-Qt} & $94.5$ & $102.1$ & $101.8$ & $\bm{92.5}$ & $170.6$ & $99.4$ & $100.7$ & $98.8$ & $101.7$ & $92.7$ & $96.9$ & $101.5$ \\ 
\bottomrule
\end{tabular*}
\begin{tablenotes}[para,flushleft]
\scriptsize{\textit{Notes}: Multi-country indicates that cross-sectional information from individual countries is used. Single-country refers to independent individual models for all countries. SV indicates the specification allowing for heteroscedastic errors, while non-SV assumes homoscedasticity. DFM -- dynamic factor model; Ham -- Hamilton's approach \citep{hamilton2017you}; HP -- Hodrick Prescott filter. LPS -- log predictive score; RMSE -- root mean squared error. \textit{1-Qt} to \textit{4-Qt} refer to the forecast horizon by quarter between one-quarter to one-year. LPS and RMSE are presented relative to independent homoscedastic univariate AR(1) processes. For LPS, the maximum value is indicated in bold, for RMSEs (in percent), the minimum is in bold, indicating the best performing specification.}
\end{tablenotes}
\end{threeparttable}
\end{center}
\end{table*}

Overall, our proposed multi-country framework DFM-SV appears to produce highly competitive out-of-sample predictions, outperforming most competing models. This finding holds true for both point and density predictive performance.  Accuracy improvements in terms of LPS tend to be substantial, irrespective of whether $g_t$ is estimated alongside the remaining model parameters and states or whether we rely on other measures of the output gap.  Considering relative RMSEs reveals that while our DFM-SV specification improves upon the benchmark model, these improvements appear to be muted and range from three percent (in the case of the one-step-ahead horizon) to 7.5 percent (for the four-quarter-ahead forecast). Only in two cases our proposed DFM-SV is slightly outperformed by multi-country versions where the latent gap component is replaced by estimates obtained  using the Hamilton (for point forecasts) and the HP  (for LPS) approach and with SV turned on. In both cases, however, DFM-SV displays the second best performance. 

Comparing the out-of-sample performance of models that utilize cross-sectional information to the ones that rely solely on aggregate EA data points towards accuracy gains of the multi-country models. Models that utilize only aggregate data generally appear to be inferior to the AR(1) model in terms of density forecasts while being slightly superior to the univariate benchmark in some cases. Specifically, the UCP model slightly improves upon the AR(1) in terms of RMSEs. These results confirm and corroborate findings in  \cite{marcellino2003macroeconomic}, who report that the inclusion of country-specific information improves out-of-sample predictions even if interest centers on predicting aggregate quantities of interest.

To sum up, Table \ref{tab:inflation-ea}  suggests that, when interest centers on forecasting euro area inflation, our proposed model framework yields strong density and point forecasts. These accuracy improvements are especially pronounced when compared to models that rely exclusively on aggregate information, highlighting the necessity to take a cross-sectional stance when forecasting inflation.

\subsection{Forecasts for individual countries}
The previous subsection provided an overall gauge on how our model performs in predicting inflation. Next, we take a cross-sectional perspective and assess whether there exist interesting cross-country differences in forecast performance.  For the sake of brevity, we focus on one-quarter-ahead forecasts in Table \ref{tab:lps-Dp-1step} and one-year-ahead predictions in Table \ref{tab:lps-Dp-4step}.  These tables include marginal LPS obtained by integrating out the remaining elements of the joint predictive density.

Starting with the one-step-ahead marginal LPS, Table \ref{tab:lps-Dp-1step} suggests that the homoscedastic variant of our proposed DFM outperforms all competing specification by large margins for most countries considered, both in terms of point and density forecasts.  Only for the Netherlands, Austria and Finland, we observe that single-country models yield  more precise density prediction whereas point forecasts for the Netherlands are most precise when single-country models are adopted.    We conjecture that this stems from the fact that these countries tend to  share a common business cycle and thus using all available cross-section information \textit{and} a single factor potentially translates into a misspecified model.

\begin{table*}[ht!]
\caption{Forecast evaluation for inflation at the one-quarter ahead forecast horizon.}\label{tab:lps-Dp-1step}\vspace*{-1.8em}
\begin{center}
\begin{threeparttable}
\scriptsize
\begin{tabular*}{\textwidth}{@{\extracolsep{\fill}} lrrrrrrrrrrrrr}
  \toprule
& \multicolumn{6}{c}{Multi-country} & \multicolumn{6}{c}{Single-country}\\
 \cmidrule(l{0pt}r{3pt}){2-7}\cmidrule(l{3pt}r{0pt}){8-13}
& \multicolumn{3}{c}{non-SV} & \multicolumn{3}{c}{SV} & \multicolumn{3}{c}{non-SV} & \multicolumn{3}{c}{SV}\\
 \cmidrule(l{0pt}r{3pt}){2-4}\cmidrule(l{3pt}r{3pt}){5-7}\cmidrule(l{3pt}r{3pt}){8-10}\cmidrule(l{3pt}r{0pt}){11-13}
& DFM & Ham & HP & DFM & Ham & HP & UCP & Ham & HP & UCP & Ham & HP \\ 
  \midrule
\textbf{LPS} &  &  &  &  &  &  &  &  &  &  &  &  \\ 
  \textit{DE} & $\bm{5.7}$ & $1.9$ & $-0.4$ & $1.1$ & $-0.4$ & $-0.4$ & $0.5$ & $4.9$ & $1.5$ & $2.2$ & $-4.7$ & $-2.2$ \\ 
  \textit{FR} & $\bm{10.6}$ & $0.3$ & $0.4$ & $6.1$ & $3.5$ & $0.5$ & $1.8$ & $2.5$ & $2.1$ & $6.7$ & $3.3$ & $6.8$ \\ 
  \textit{IT} & $\bm{3.4}$ & $-3.1$ & $-1.9$ & $0.9$ & $1.3$ & $-2.0$ & $-2.3$ & $-6.8$ & $-2.7$ & $-7.8$ & $-8.1$ & $-6.4$ \\ 
  \textit{ES} & $\bm{2.6}$ & $-10.2$ & $-10.4$ & $0.3$ & $-4.2$ & $-7.6$ & $-8.6$ & $-1.4$ & $-2.6$ & $-6.0$ & $-7.2$ & $-9.7$ \\ 
  \textit{NL} & $0.7$ & $-6.5$ & $-5.5$ & $-1.2$ & $0.1$ & $-0.8$ & $-4.3$ & $-0.3$ & $\bm{1.7}$ & $-4.7$ & $-12.3$ & $-12.6$ \\ 
  \textit{BE} & $\bm{10.4}$ & $1.5$ & $2.5$ & $6.9$ & $4.0$ & $0.9$ & $-0.8$ & $-1.1$ & $-0.4$ & $4.7$ & $4.6$ & $4.5$ \\ 
  \textit{AT} & $6.5$ & $-0.5$ & $-6.0$ & $3.2$ & $-2.5$ & $-5.0$ & $1.4$ & $-0.6$ & $1.6$ & $\bm{7.6}$ & $0.7$ & $0.9$ \\ 
  \textit{FI} & $5.0$ & $3.2$ & $3.9$ & $4.6$ & $3.8$ & $3.7$ & $3.7$ & $4.6$ & $\bm{5.1}$ & $2.3$ & $-0.3$ & $0.6$ \\ 
  \textit{PT} & $\bm{3.4}$ & $-4.0$ & $-3.8$ & $0.8$ & $-2.5$ & $-3.0$ & $2.2$ & $-0.5$ & $0.4$ & $0.5$ & $-5.8$ & $-2.9$ \\ 
  \textit{GR} & $\bm{3.9}$ & $-7.0$ & $-7.5$ & $-2.8$ & $-8.8$ & $-8.7$ & $0.6$ & $-3.5$ & $-4.9$ & $-2.3$ & $-9.4$ & $-11.1$ \\ 
  \midrule
  \textbf{RMSE} &  &  &  &  &  &  &  &  &  &  &  &  \\ 
  \textit{DE} & $\bm{91.3}$ & $101.1$ & $104.9$ & $100.6$ & $113.9$ & $127.5$ & $104.3$ & $95.2$ & $98.7$ & $100.3$ & $115.2$ & $111.3$ \\ 
  \textit{FR} & $\bm{81.7}$ & $102.5$ & $106.3$ & $95.9$ & $105.2$ & $131.1$ & $99.0$ & $97.2$ & $100.8$ & $99.3$ & $118.3$ & $113.3$ \\ 
  \textit{IT} & $\bm{88.9}$ & $102.0$ & $113.1$ & $97.4$ & $101.2$ & $110.0$ & $97.2$ & $100.4$ & $100.8$ & $103.3$ & $107.9$ & $106.7$ \\ 
  \textit{ES} & $\bm{86.8}$ & $107.0$ & $110.9$ & $95.5$ & $104.5$ & $122.3$ & $108.6$ & $100.0$ & $100.8$ & $107.7$ & $119.9$ & $125.8$ \\ 
  \textit{NL} & $100.1$ & $105.4$ & $103.2$ & $100.0$ & $99.6$ & $112.0$ & $103.8$ & $101.9$ & $\bm{99.4}$ & $101.4$ & $115.1$ & $112.5$ \\ 
  \textit{BE} & $\bm{78.0}$ & $100.0$ & $112.1$ & $90.6$ & $101.9$ & $131.2$ & $99.2$ & $100.5$ & $99.2$ & $98.0$ & $105.0$ & $105.7$ \\ 
  \textit{AT} & $\bm{84.2}$ & $101.7$ & $103.4$ & $92.7$ & $102.2$ & $115.4$ & $97.2$ & $98.2$ & $101.4$ & $87.5$ & $108.8$ & $110.1$ \\ 
  \textit{FI} & $\bm{94.3}$ & $99.5$ & $100.7$ & $100.3$ & $101.8$ & $111.9$ & $99.0$ & $96.6$ & $98.5$ & $100.0$ & $129.1$ & $108.9$ \\ 
  \textit{PT} & $\bm{89.2}$ & $103.0$ & $100.9$ & $95.7$ & $104.0$ & $108.1$ & $98.8$ & $101.1$ & $102.0$ & $97.2$ & $130.7$ & $116.9$ \\ 
  \textit{GR} & $\bm{86.0}$ & $102.5$ & $104.7$ & $95.8$ & $109.7$ & $123.8$ & $97.4$ & $104.7$ & $109.3$ & $98.4$ & $110.0$ & $111.4$ \\ 
   \bottomrule
\end{tabular*}
\begin{tablenotes}[para,flushleft]
\scriptsize{\textit{Notes}: Multi-country indicates that cross-sectional information from individual countries is used. Single-country refers to independent individual models for all countries. SV indicates the specification allowing for heteroscedastic errors, while non-SV assumes homoscedasticity. DFM -- dynamic factor model; Ham -- Hamilton's approach \citep{hamilton2017you}; HP -- Hodrick Prescott filter. LPS -- log predictive score; RMSE -- root mean squared error. LPS and RMSE are presented relative to independent homoscedastic univariate AR(1) processes. For LPS, the maximum value is indicated in bold, for RMSEs (in percent), the minimum is in bold, indicating the best performing specification.}
\end{tablenotes}
\end{threeparttable}
\end{center}
\end{table*}

Considering accuracy gains from controlling for heteroscedasticy shows that for most countries, explicitly allowing for SV translates into weaker point and density forecasts relative to the homoscedastic counterparts. This result is in contrast to the findings based on using the full predictive distribution of inflation reported in Table \ref{tab:inflation-ea} and the literature on inflation forecasting \citep{stock2007has,stella2013state,jarocinski2018inflation}. The reasons for this slightly inferior performance of SV specifications in terms of marginal LPS are at least threefold. First, the hold-out period covers the beginning of the global financial crisis, implying that error variances are already tilted upwards. Second, our DFM-SV specification constitutes a parsimonious means of modeling a large dimensional time-varying variance-covariance matrix. Thus, if interest centers on capturing the potentially time-varying nature of covariances (which is relevant if the full predictive density is evaluated), predictive gains in terms of density forecasts tend to increase with the dimension of the model \citep{KASTNER201998}. Third, and contrasting accuracy differences between models that treat the gap component as observed as opposed to latent, we generally find that multi-country models profit from explicitly controlling for estimation uncertainty surrounding $g_t$. This premium in predictive accuracy stems from the fact that integrating out $g_t$ from the predictive density translates into a heavy-tailed marginal predictive distribution that is capable of handling outlying values well. This lowers the necessity to explicitly control for stochastic volatility, especially for data at quarterly frequency.

\begin{table*}[ht!]
\caption{Forecast evaluation for inflation at the one-year-ahead forecast horizon.}\label{tab:lps-Dp-4step}\vspace*{-1.8em}
\begin{center}
\begin{threeparttable}
\scriptsize
\begin{tabular*}{\textwidth}{@{\extracolsep{\fill}} lrrrrrrrrrrrrr}
  \toprule
& \multicolumn{6}{c}{Multi-country} & \multicolumn{6}{c}{Single-country}\\
 \cmidrule(l{0pt}r{3pt}){2-7}\cmidrule(l{3pt}r{0pt}){8-13}
& \multicolumn{3}{c}{non-SV} & \multicolumn{3}{c}{SV} & \multicolumn{3}{c}{non-SV} & \multicolumn{3}{c}{SV}\\
 \cmidrule(l{0pt}r{3pt}){2-4}\cmidrule(l{3pt}r{3pt}){5-7}\cmidrule(l{3pt}r{3pt}){8-10}\cmidrule(l{3pt}r{0pt}){11-13}
& DFM & Ham & HP & DFM & Ham & HP & UCP & Ham & HP & UCP & Ham & HP \\ 
  \midrule
\textbf{LPS} &  &  &  &  &  &  &  &  &  &  &  &  \\ 
  \textit{DE} & $3.6$ & $-0.7$ & $-1.0$ & $1.7$ & $-1.4$ & $-2.0$ & $-2.6$ & $\bm{5.6}$ & $0.5$ & $0.3$ & $0.1$ & $-0.6$ \\ 
  \textit{FR} & $\bm{3.5}$ & $-7.2$ & $-9.2$ & $0.2$ & $-4.6$ & $-9.4$ & $-4.6$ & $-4.5$ & $-0.9$ & $-3.1$ & $-2.5$ & $-3.3$ \\ 
  \textit{IT} & $-6.1$ & $-7.7$ & $-9.9$ & $\bm{-1.9}$ & $-5.8$ & $-10.2$ & $-14.7$ & $-23.5$ & $-14.8$ & $-22.1$ & $-24.3$ & $-24.8$ \\ 
  \textit{ES} & $\bm{1.2}$ & $-16.7$ & $-19.3$ & $0.8$ & $-9.5$ & $-16.0$ & $-33.7$ & $-2.7$ & $-4.0$ & $-11.8$ & $-12.6$ & $-20.7$ \\ 
  \textit{NL} & $-1.4$ & $-7.6$ & $-9.4$ & $\bm{1.0}$ & $-1.1$ & $-0.7$ & $-7.4$ & $-1.8$ & $-0.7$ & $0.3$ & $-18.4$ & $-14.8$ \\ 
  \textit{BE} & $\bm{9.1}$ & $-6.2$ & $-7.4$ & $4.4$ & $1.3$ & $-6.9$ & $-3.5$ & $-7.0$ & $-4.9$ & $2.2$ & $4.8$ & $-1.7$ \\ 
  \textit{AT} & $1.7$ & $-8.5$ & $-10.5$ & $-0.7$ & $-8.9$ & $-6.6$ & $-5.0$ & $-5.1$ & $-3.5$ & $\bm{5.1}$ & $2.2$ & $-5.4$ \\ 
  \textit{FI} & $-6.0$ & $-6.4$ & $-5.6$ & $-3.8$ & $-3.4$ & $-3.2$ & $1.0$ & $-0.8$ & $1.2$ & $-6.1$ & $\bm{5.9}$ & $-2.7$ \\ 
  \textit{PT} & $\bm{1.9}$ & $-7.7$ & $-10.3$ & $-3.1$ & $-10.4$ & $-7.7$ & $-0.6$ & $-5.3$ & $-2.5$ & $-3.0$ & $-1.4$ & $-8.3$ \\ 
  \textit{GR} & $\bm{9.3}$ & $-13.4$ & $-16.3$ & $-2.1$ & $-12.8$ & $-19.7$ & $-3.5$ & $-6.0$ & $-14.2$ & $-0.8$ & $-14.0$ & $-21.3$ \\ 
  \midrule
  \textbf{RMSE} &  &  &  &  &  &  &  &  &  &  &  &  \\ 
  \textit{DE} & $92.0$ & $102.0$ & $106.1$ & $96.4$ & $104.8$ & $102.8$ & $114.4$ & $\bm{89.4}$ & $99.1$ & $98.0$ & $95.5$ & $97.7$ \\ 
  \textit{FR} & $\bm{85.9}$ & $107.2$ & $110.7$ & $94.3$ & $107.8$ & $111.6$ & $103.9$ & $100.1$ & $102.0$ & $99.5$ & $105.3$ & $102.4$ \\ 
  \textit{IT} & $\bm{93.1}$ & $103.2$ & $113.2$ & $94.8$ & $101.8$ & $98.4$ & $95.2$ & $101.7$ & $101.0$ & $101.5$ & $98.5$ & $100.3$ \\ 
  \textit{ES} & $\bm{88.1}$ & $110.5$ & $116.6$ & $94.8$ & $108.7$ & $110.5$ & $141.7$ & $97.7$ & $99.8$ & $96.7$ & $103.2$ & $103.6$ \\ 
  \textit{NL} & $99.6$ & $107.3$ & $124.3$ & $96.1$ & $\bm{93.6}$ & $95.6$ & $116.8$ & $102.7$ & $99.8$ & $93.6$ & $98.7$ & $98.9$ \\ 
  \textit{BE} & $\bm{72.5}$ & $104.8$ & $115.2$ & $81.4$ & $99.1$ & $105.9$ & $95.5$ & $96.7$ & $96.4$ & $95.0$ & $93.2$ & $95.0$ \\ 
  \textit{AT} & $\bm{82.6}$ & $98.9$ & $101.2$ & $87.9$ & $101.0$ & $96.4$ & $97.7$ & $97.0$ & $104.1$ & $83.9$ & $91.5$ & $100.2$ \\ 
  \textit{FI} & $97.5$ & $97.4$ & $102.9$ & $97.1$ & $98.6$ & $98.4$ & $95.8$ & $95.2$ & $93.8$ & $102.1$ & $\bm{81.2}$ & $95.5$ \\ 
  \textit{PT} & $\bm{84.0}$ & $101.6$ & $102.4$ & $88.7$ & $106.3$ & $101.2$ & $101.0$ & $105.5$ & $103.9$ & $95.7$ & $103.1$ & $103.7$ \\ 
  \textit{GR} & $\bm{78.5}$ & $107.3$ & $109.0$ & $91.0$ & $109.8$ & $116.0$ & $103.5$ & $108.6$ & $123.4$ & $99.0$ & $110.5$ & $120.0$ \\ 
   \bottomrule
\end{tabular*}
\begin{tablenotes}[para,flushleft]
\scriptsize{\textit{Notes}: Multi-country indicates that cross-sectional information from individual countries is used. Single-country refers to independent individual models for all countries. SV indicates the specification allowing for heteroscedastic errors, while non-SV assumes homoscedasticity. DFM -- dynamic factor model; Ham -- Hamilton's approach \citep{hamilton2017you}; HP -- Hodrick Prescott filter. LPS -- log predictive score; RMSE -- root mean squared error. LPS and RMSE are presented relative to independent homoscedastic univariate AR(1) processes. For LPS, the maximum value is indicated in bold, for RMSEs (in percent), the minimum is in bold, indicating the best performing specification.}
\end{tablenotes}
\end{threeparttable}
\end{center}
\end{table*}

Turning attention to the one-year-ahead forecast horizon, Table \ref{tab:lps-Dp-4step} shows similar results to those reported for the one-quarter-ahead horizon. For this longer forecast horizon, the homoscedastic DFM setup appears to be particularly successful in terms of producing accurate point predictions, which is not surprising given the fact that for higher-order forecasts, the log-volatilities approach their stationary distribution. The  predictive performance  in terms of point forecasts of the  DFM  model is comparable to its heteroscedastic counterpart. Unlike the remaining alternative models, both DFM and DFM-SV also manage to notably outperform the AR(1) benchmark for the one-year-ahead horizon. In terms of density forecasts, however, the predictive dominance of DFM appears less distinctive.

An overall comparison between multi-country and single-country models for the one-year-ahead horizon again  reveals no clear pattern. However, this is particularly due to the strong performance of our proposed multi-country frameworks DFM and DFM-SV. Without these two specifications, Table \ref{tab:lps-Dp-4step} shows that single-country models appear to be preferable compared to the multi-country setups. However, for one-year-ahead predictions, the table again highlights the importance for  including  cross-sectional information to produce accurate point forecasts for inflation  in Portugal and Greece. 

\section{Concluding remarks}
In this paper, we develop a multivariate Bayesian dynamic factor model with stochastic volatility for analyzing euro area business cycles. The multi-country framework decomposes country-specific output and inflation series into idiosyncratic non-stationary trends and a joint stationary cyclical component. This enables us to exploit cross-sectional information and obtain an EA-wide measure of the  output gap used for structural analysis and inflation forecasting. A key model feature is to allow for heteroscedastic error terms and comovements in the trends using a flexible factor stochastic volatility structure. The setup is completed by considering time variation also in the variances of the measurement equations. The proposed Bayesian model alleviates concerns of overparameterization via global-local shrinkage priors that push the model towards a homoscedastic specification, but allows for time-varying variances if necessary. 

In an empirical section, we study both in-sample features and out-of-sample predictive performance of the proposed model. We compare the obtained measure of the output gap to a set of competing approaches for estimation and discuss the role of time variation in error variances. The analysis is complemented by an empirical assessment regarding the slope of the Philips curve across EA member states. In a forecasting exercise, the paper provides evidence that accounting for a common euro area output gap component produces competitive forecasts for inflation both on the aggregate EA, but also the country level.

\small{\scfont\setstretch{0.85}
\addcontentsline{toc}{section}{References}
\bibliographystyle{custom}
\bibliography{dfm}}

\clearpage
\begin{appendices}\crefalias{section}{appsec}
\setcounter{equation}{0}
\renewcommand\theequation{A.\arabic{equation}}

\section{Full conditional posterior distributions}\label{sec:MCMC algorithm}
It is worth noting that the joint posterior distribution of the model parameters and the set of latent states is intractable. Fortunately, the full conditional posterior distributions for most quantities are of a simple form and thus amenable to standard Gibbs updating. In order to obtain a draw from the joint posterior we design a straightforward Markov chain Monte Carlo (MCMC) algorithm that cycles through the following steps:
\begin{enumerate}[label=(\roman*),align=left]
\item Simulate the full history of $\{\bm{f}_t\}_{t=1}^T$ using a forward filtering backward sampling algorithm (\citealt{carter1994gibbs}, \citealt{fruhwirth1994applied}).
\item Draw the sequence of log-volatilities $\{h_{kit}\}_{t=1}^T, \{\upsilon_{lt}\}_{t=1}^T, \{\omega_{rt}\}_{t=1}^T,$ for all $i, k, l, r$ as well as the parameters in the corresponding state equations independently using the algorithm proposed in \cite{kastner2014ancillarity}. 
\item Conditional on the unobserved components, we simulate the loadings $\alpha_{ki}$ and $\beta_{ki}$ by estimating $2N$ independent regression models with heteroscedastic innovations
\item Conditional on the other parameters of the model, we simulate the history of the factors $\{\bm{z}_t\}_{t=1}^{T}$ driving the covariances between the country-specific trend components based on the regressions and quantities given in \autoref{eq:etayt} and \autoref{eq:etapit}.
\item The free elements in $\bm{\Lambda}$ conditionally on knowing the full history of the factors $\bm{z}_t$ can be sampled on an equation-by-equation basis involving a sequence of standard linear regression models with heteroscedastic errors \citep[see also][]{aguilar2000bayesian,KASTNER201998}.
\item The parameters $Q$ and $\gamma$ are updated in a block by using a standard random walk Metropolis Hastings algorithm.
\item Sample $\xi_h$, $\xi_\omega$ and $\xi_\upsilon$ from a Gamma distributed conditional posterior distribution.
\end{enumerate}
Steps (i) to (v) are standard and easily executed. Steps (vi) and (vii) deserve more attention. The full conditional posterior distributions of $B_{hki}$, $B_{\omega r}$, and $B_{\upsilon l}$ are similar, and we thus only present specifics for one of them, $B_{\omega r}$. The conditional posterior of this parameter follows a GIG distribution that is obtained by combining the conditional density $p(\sqrt{\vartheta_{\omega r}}| B_{\omega r})$ with the conditional prior $p(B_{\omega r}| \xi_\omega)$,
\begin{equation}
B_{\omega_r}|\bullet \sim \mathcal{GIG}(\kappa_\omega - 1/2, \vartheta_{\omega r}, \xi_\omega \kappa_\omega),
\end{equation}
where $\bullet$ denotes conditioning on all remaining quantities of the model.

To obtain the full conditional posterior distribution for the global scaling parameters that is again similar for $\xi_h$, $\xi_\omega$ and $\xi_\upsilon$, we combine the joint density $\prod_{r=1}^M p(B_{\omega r}|\xi_\omega)$ with the prior $p(\xi_\omega)$. This yields a Gamma distributed conditional posterior distribution,
\begin{equation}
\xi_\omega | \bullet \sim \mathcal{G}\left(c_0+\kappa_\omega M, c_1 + \frac{\kappa_\omega}{2} \sum_{r=1}^M B_{\omega r}\right).
\end{equation}
This setup completes the full simulation-based algorithm. We iterate the steps above for $50,000$ times with a burn-in period of the first $25,000$ cycles. The obtained results provide evidence for satisfactory convergence properties of the MCMC algorithm.

\ifx
\section{Forecasting results for output}
\begin{table*}[ht]
\caption{Forecast evaluation for euro area output.}\label{tab:output-ea}\vspace*{-1.8em}
\begin{center}
\begin{threeparttable}
\scriptsize
\begin{tabular*}{\textwidth}{@{\extracolsep{\fill}} lrrrrrrrrrrrrrr}
  \toprule
& \multicolumn{6}{c}{Multi-country} & \multicolumn{6}{c}{Single-country}\\
 \cmidrule(l{0pt}r{3pt}){2-7}\cmidrule(l{3pt}r{0pt}){8-13}
& \multicolumn{3}{c}{non-SV} & \multicolumn{3}{c}{SV} & \multicolumn{3}{c}{non-SV} & \multicolumn{3}{c}{SV}\\
 \cmidrule(l{0pt}r{3pt}){2-4}\cmidrule(l{3pt}r{3pt}){5-7}\cmidrule(l{3pt}r{3pt}){8-10}\cmidrule(l{3pt}r{0pt}){11-13}
& DFM & Ham & HP & DFM & Ham & HP & UCP & Ham & HP & UCP & Ham & HP \\ 
\midrule
\textbf{LPS} &  &  &  &  &  &  &  &  &  &  &  &  \\ 
  \textit{1-Qt} & $715.5$ & $732.7$ & $744.5$ & $740.6$ & $730.0$ & $\bm{752.0}$ & $696.0$ & $715.7$ & $713.5$ & $687.0$ & $717.7$ & $710.9$ \\ 
  \textit{2-Qt} & $1347.6$ & $1371.6$ & $1379.4$ & $1370.2$ & $1365.3$ & $\bm{1384.1}$ & $1297.4$ & $1320.8$ & $1282.7$ & $1267.4$ & $1335.2$ & $1269.1$ \\ 
  \textit{3-Qt} & $1800.0$ & $1825.5$ & $1827.5$ & $1819.4$ & $1812.7$ & $\bm{1831.8}$ & $1740.5$ & $1757.5$ & $1673.5$ & $1706.5$ & $1759.0$ & $1651.5$ \\ 
  \textit{4-Qt} & $2232.6$ & $\bm{2261.6}$ & $2257.3$ & $2250.8$ & $2244.1$ & $2257.9$ & $2169.1$ & $2180.9$ & $2051.6$ & $2133.2$ & $2164.8$ & $2017.6$ \\ 
  \midrule
  \textbf{RMSE} &  &  &  &  &  &  &  &  &  &  &  &  \\ 
  \textit{1-Qt} & $104.9$ & $85.7$ & $75.8$ & $89.3$ & $93.2$ & $\bm{72.0}$ & $83.2$ & $89.2$ & $79.1$ & $87.2$ & $86.1$ & $79.0$ \\ 
  \textit{2-Qt} & $114.3$ & $85.3$ & $78.5$ & $95.0$ & $89.2$ & $\bm{71.7}$ & $88.9$ & $84.8$ & $81.5$ & $90.2$ & $83.4$ & $82.5$ \\ 
  \textit{3-Qt} & $123.0$ & $85.0$ & $79.6$ & $103.0$ & $136.4$ & $\bm{72.2}$ & $92.4$ & $81.6$ & $81.3$ & $91.4$ & $84.3$ & $82.2$ \\ 
  \textit{4-Qt} & $134.4$ & $81.9$ & $77.4$ & $111.9$ & $219.2$ & $\bm{71.5}$ & $97.5$ & $77.0$ & $79.2$ & $93.9$ & $85.3$ & $79.9$ \\ 
\bottomrule
\end{tabular*}
\begin{tablenotes}[para,flushleft]
\scriptsize{\textit{Notes}: Multi-country indicates that cross-sectional information from individual countries is used. Single-country refers to independent individual models for all countries. SV indicates the specification allowing for heteroscedastic errors, while non-SV assumes homoscedasticity. DFM -- dynamic factor model; Ham -- Hamilton's approach \citep{hamilton2017you}; HP -- Hodrick Prescott filter. LPS -- log predictive score; RMSE -- root mean squared error. \textit{1-Qt} to \textit{4-Qt} refer to the forecast horizon by quarter between one-quarter to one-year. LPS and RMSE are presented relative to independent homoscedastic univariate AR(1) processes. For LPS, the maximum value is indicated in bold, for RMSEs (in percent), the minimum is in bold, indicating the best performing specification.}
\end{tablenotes}
\end{threeparttable}
\end{center}
\end{table*}

\begin{table*}[ht]
\caption{Forecast evaluation for output at the one-quarter ahead forecast horizon.}\label{tab:lps-y-1step}\vspace*{-1.8em}
\begin{center}
\begin{threeparttable}
\scriptsize
\begin{tabular*}{\textwidth}{@{\extracolsep{\fill}} lrrrrrrrrrrrrr}
  \toprule
& \multicolumn{6}{c}{Multi-country} & \multicolumn{6}{c}{Single-country}\\
 \cmidrule(l{0pt}r{3pt}){2-7}\cmidrule(l{3pt}r{0pt}){8-13}
& \multicolumn{3}{c}{non-SV} & \multicolumn{3}{c}{SV} & \multicolumn{3}{c}{non-SV} & \multicolumn{3}{c}{SV}\\
 \cmidrule(l{0pt}r{3pt}){2-4}\cmidrule(l{3pt}r{3pt}){5-7}\cmidrule(l{3pt}r{3pt}){8-10}\cmidrule(l{3pt}r{0pt}){11-13}
& DFM & Ham & HP & DFM & Ham & HP & UCP & Ham & HP & UCP & Ham & HP \\ 
  \midrule
\textbf{LPS} &  &  &  &  &  &  &  &  &  &  &  &  \\ 
  \textit{DE} & $47.8$ & $62.5$ & $64.3$ & $51.0$ & $66.0$ & $\bm{69.0}$ & $50.0$ & $51.4$ & $57.8$ & $29.5$ & $53.0$ & $56.3$ \\ 
  \textit{FR} & $30.3$ & $31.5$ & $31.7$ & $29.9$ & $33.0$ & $33.3$ & $25.6$ & $19.0$ & $30.4$ & $29.2$ & $17.5$ & $\bm{36.4}$ \\ 
  \textit{IT} & $39.3$ & $42.7$ & $48.3$ & $41.6$ & $44.4$ & $\bm{50.9}$ & $42.1$ & $42.6$ & $48.0$ & $44.2$ & $45.7$ & $46.9$ \\ 
  \textit{ES} & $100.5$ & $99.3$ & $96.6$ & $99.5$ & $99.3$ & $97.0$ & $100.6$ & $83.1$ & $105.4$ & $99.8$ & $84.3$ & $\bm{109.5}$ \\ 
  \textit{NL} & $56.5$ & $55.2$ & $55.4$ & $54.8$ & $56.3$ & $\bm{57.0}$ & $34.9$ & $45.9$ & $53.6$ & $32.9$ & $46.9$ & $56.8$ \\ 
  \textit{BE} & $24.3$ & $24.5$ & $24.5$ & $23.7$ & $25.5$ & $26.7$ & $17.4$ & $21.7$ & $29.0$ & $20.6$ & $19.8$ & $\bm{31.6}$ \\ 
  \textit{AT} & $46.0$ & $47.5$ & $47.7$ & $46.0$ & $48.9$ & $49.4$ & $45.3$ & $38.1$ & $51.6$ & $46.9$ & $41.4$ & $\bm{53.1}$ \\ 
  \textit{FI} & $138.7$ & $138.3$ & $142.9$ & $137.0$ & $\bm{145.4}$ & $142.6$ & $133.9$ & $135.4$ & $137.6$ & $127.8$ & $130.7$ & $130.9$ \\ 
  \textit{PT} & $44.4$ & $43.8$ & $44.1$ & $43.2$ & $44.2$ & $\bm{44.7}$ & $43.6$ & $39.5$ & $39.5$ & $35.4$ & $42.2$ & $35.3$ \\ 
  \textit{GR} & $158.0$ & $158.6$ & $156.0$ & $153.7$ & $156.8$ & $155.0$ & $\bm{172.1}$ & $144.3$ & $136.8$ & $157.5$ & $86.7$ & $85.1$ \\ 
  \midrule
  \textbf{RMSE} &  &  &  &  &  &  &  &  &  &  &  &  \\ 
  \textit{DE} & $111.7$ & $91.1$ & $88.8$ & $107.6$ & $88.1$ & $\bm{85.0}$ & $106.4$ & $121.9$ & $101.5$ & $133.7$ & $121.1$ & $98.2$ \\ 
  \textit{FR} & $87.7$ & $84.4$ & $83.1$ & $88.7$ & $78.8$ & $80.0$ & $90.8$ & $103.0$ & $78.0$ & $87.5$ & $121.1$ & $\bm{75.8}$ \\ 
  \textit{IT} & $91.8$ & $88.6$ & $76.4$ & $91.6$ & $86.2$ & $\bm{73.9}$ & $89.6$ & $78.0$ & $83.9$ & $92.2$ & $76.9$ & $84.8$ \\ 
  \textit{ES} & $68.8$ & $71.2$ & $76.2$ & $70.1$ & $70.0$ & $74.4$ & $67.9$ & $114.8$ & $49.4$ & $71.6$ & $113.7$ & $\bm{49.1}$ \\ 
  \textit{NL} & $\bm{87.9}$ & $90.1$ & $89.0$ & $89.7$ & $88.6$ & $88.6$ & $116.5$ & $112.1$ & $95.1$ & $128.2$ & $120.0$ & $88.9$ \\ 
  \textit{BE} & $90.8$ & $90.0$ & $90.3$ & $92.8$ & $85.1$ & $85.5$ & $101.2$ & $79.8$ & $74.3$ & $96.0$ & $96.7$ & $\bm{72.0}$ \\ 
  \textit{AT} & $92.9$ & $90.4$ & $91.1$ & $94.0$ & $87.1$ & $87.5$ & $92.8$ & $101.2$ & $80.5$ & $92.1$ & $105.2$ & $\bm{78.9}$ \\ 
  \textit{FI} & $88.1$ & $87.9$ & $86.7$ & $88.7$ & $\bm{83.5}$ & $85.3$ & $103.8$ & $99.0$ & $94.2$ & $108.1$ & $98.8$ & $94.9$ \\ 
  \textit{PT} & $84.9$ & $87.6$ & $85.3$ & $87.2$ & $85.5$ & $\bm{84.6}$ & $85.4$ & $91.8$ & $96.6$ & $100.1$ & $86.6$ & $96.2$ \\ 
  \textit{GR} & $104.5$ & $103.6$ & $106.3$ & $106.8$ & $108.6$ & $109.2$ & $\bm{86.2}$ & $175.2$ & $124.7$ & $95.6$ & $173.4$ & $124.7$ \\ 
   \bottomrule
\end{tabular*}
\begin{tablenotes}[para,flushleft]
\scriptsize{\textit{Notes}: Multi-country indicates that cross-sectional information from individual countries is used. Single-country refers to independent individual models for all countries. SV indicates the specification allowing for heteroscedastic errors, while non-SV assumes homoscedasticity. DFM -- dynamic factor model; Ham -- Hamilton's approach \citep{hamilton2017you}; HP -- Hodrick Prescott filter. LPS -- log predictive score; RMSE -- root mean squared error. LPS and RMSE are presented relative to independent homoscedastic univariate AR(1) processes. For LPS, the maximum value is indicated in bold, for RMSEs (in percent), the minimum is in bold, indicating the best performing specification.}
\end{tablenotes}
\end{threeparttable}
\end{center}
\end{table*}

\begin{table*}[ht]
\caption{Forecast evaluation for output at the one-year ahead forecast horizon.}\label{tab:lps-y-4step}\vspace*{-1.8em}
\begin{center}
\begin{threeparttable}
\scriptsize
\begin{tabular*}{\textwidth}{@{\extracolsep{\fill}} lrrrrrrrrrrrrr}
  \toprule
& \multicolumn{6}{c}{Multi-country} & \multicolumn{6}{c}{Single-country}\\
 \cmidrule(l{0pt}r{3pt}){2-7}\cmidrule(l{3pt}r{0pt}){8-13}
& \multicolumn{3}{c}{non-SV} & \multicolumn{3}{c}{SV} & \multicolumn{3}{c}{non-SV} & \multicolumn{3}{c}{SV}\\
 \cmidrule(l{0pt}r{3pt}){2-4}\cmidrule(l{3pt}r{3pt}){5-7}\cmidrule(l{3pt}r{3pt}){8-10}\cmidrule(l{3pt}r{0pt}){11-13}
& DFM & Ham & HP & DFM & Ham & HP & UCP & Ham & HP & UCP & Ham & HP \\ 
  \midrule
\textbf{LPS} &  &  &  &  &  &  &  &  &  &  &  &  \\ 
  \textit{DE} & $143.2$ & $166.1$ & $145.8$ & $133.2$ & $\bm{172.0}$ & $142.1$ & $122.6$ & $153.3$ & $140.6$ & $103.9$ & $142.3$ & $130.9$ \\ 
  \textit{FR} & $130.2$ & $130.8$ & $131.4$ & $130.5$ & $133.6$ & $131.3$ & $129.4$ & $131.9$ & $\bm{136.1}$ & $126.9$ & $122.4$ & $134.3$ \\ 
  \textit{IT} & $159.5$ & $163.0$ & $151.9$ & $165.4$ & $169.7$ & $153.2$ & $155.0$ & $176.2$ & $145.7$ & $149.8$ & $\bm{182.4}$ & $123.9$ \\ 
  \textit{ES} & $524.6$ & $519.9$ & $\bm{527.8}$ & $524.1$ & $524.1$ & $523.2$ & $466.6$ & $508.5$ & $520.5$ & $517.1$ & $477.5$ & $485.3$ \\ 
  \textit{NL} & $162.8$ & $161.2$ & $159.9$ & $161.2$ & $\bm{163.0}$ & $160.3$ & $126.3$ & $159.3$ & $160.1$ & $134.0$ & $158.4$ & $148.2$ \\ 
  \textit{BE} & $101.5$ & $102.3$ & $101.2$ & $101.4$ & $106.3$ & $103.7$ & $98.5$ & $\bm{108.4}$ & $107.9$ & $82.2$ & $96.7$ & $107.3$ \\ 
  \textit{AT} & $170.5$ & $170.6$ & $167.5$ & $168.4$ & $173.4$ & $169.3$ & $171.7$ & $\bm{179.8}$ & $178.7$ & $171.3$ & $176.7$ & $177.5$ \\ 
  \textit{FI} & $358.5$ & $361.3$ & $360.3$ & $358.0$ & $\bm{367.7}$ & $360.6$ & $350.8$ & $365.6$ & $347.4$ & $323.0$ & $360.1$ & $331.9$ \\ 
  \textit{PT} & $\bm{162.0}$ & $158.5$ & $161.1$ & $161.5$ & $159.7$ & $158.6$ & $156.5$ & $156.8$ & $138.2$ & $151.2$ & $147.2$ & $85.1$ \\ 
  \textit{GR} & $472.0$ & $476.8$ & $475.9$ & $476.5$ & $475.8$ & $473.7$ & $\bm{529.0}$ & $451.0$ & $343.7$ & $500.5$ & $363.3$ & $161.4$ \\
  \midrule 
  \textbf{RMSE} &  &  &  &  &  &  &  &  &  &  &  &  \\ 
  \textit{DE} & $106.7$ & $82.3$ & $99.1$ & $106.6$ & $\bm{78.2}$ & $90.4$ & $155.0$ & $117.9$ & $95.4$ & $169.1$ & $117.2$ & $94.2$ \\ 
  \textit{FR} & $83.2$ & $84.6$ & $81.6$ & $83.4$ & $79.0$ & $79.4$ & $97.2$ & $88.7$ & $78.6$ & $97.1$ & $115.8$ & $\bm{77.7}$ \\ 
  \textit{IT} & $85.0$ & $85.3$ & $87.2$ & $83.8$ & $85.9$ & $\bm{83.7}$ & $109.8$ & $91.5$ & $94.9$ & $112.9$ & $89.0$ & $96.2$ \\ 
  \textit{ES} & $\bm{64.5}$ & $67.8$ & $64.6$ & $65.0$ & $67.8$ & $65.4$ & $148.9$ & $108.5$ & $74.9$ & $79.6$ & $137.2$ & $75.6$ \\ 
  \textit{NL} & $\bm{75.1}$ & $79.2$ & $78.0$ & $76.5$ & $77.0$ & $75.8$ & $199.5$ & $89.5$ & $81.1$ & $184.6$ & $87.9$ & $78.9$ \\ 
  \textit{BE} & $85.5$ & $85.9$ & $86.4$ & $86.5$ & $80.7$ & $81.9$ & $115.4$ & $\bm{71.0}$ & $74.6$ & $152.1$ & $105.1$ & $73.2$ \\ 
  \textit{AT} & $84.5$ & $86.2$ & $86.4$ & $86.2$ & $83.0$ & $82.6$ & $89.8$ & $79.1$ & $77.7$ & $84.0$ & $91.7$ & $\bm{76.3}$ \\ 
  \textit{FI} & $71.8$ & $72.0$ & $72.5$ & $72.8$ & $71.3$ & $71.0$ & $132.8$ & $\bm{68.0}$ & $78.0$ & $162.3$ & $75.6$ & $78.7$ \\ 
  \textit{PT} & $\bm{76.3}$ & $80.4$ & $77.3$ & $77.4$ & $80.1$ & $77.7$ & $100.8$ & $90.3$ & $97.7$ & $102.1$ & $86.3$ & $99.0$ \\ 
  \textit{GR} & $82.7$ & $83.5$ & $83.3$ & $84.4$ & $85.7$ & $84.7$ & $\bm{71.1}$ & $138.1$ & $122.2$ & $83.1$ & $137.3$ & $121.7$ \\ 
   \bottomrule
\end{tabular*}
\begin{tablenotes}[para,flushleft]
\scriptsize{\textit{Notes}: Multi-country indicates that cross-sectional information from individual countries is used. Single-country refers to independent individual models for all countries. SV indicates the specification allowing for heteroscedastic errors, while non-SV assumes homoscedasticity. DFM -- dynamic factor model; Ham -- Hamilton's approach \citep{hamilton2017you}; HP -- Hodrick Prescott filter. LPS -- log predictive score; RMSE -- root mean squared error. LPS and RMSE are presented relative to independent homoscedastic univariate AR(1) processes. For LPS, the maximum value is indicated in bold, for RMSEs (in percent), the minimum is in bold, indicating the best performing specification.}
\end{tablenotes}
\end{threeparttable}
\end{center}
\end{table*}
\fi

\end{appendices}
\end{document}